\documentclass[journal]{IEEEtran}
\usepackage{epsfig,graphicx,subfigure,psfrag,amsmath,cases}
\usepackage{latexsym,amssymb,amsmath,epsfig,subfigure,algorithm,mathtools}
\usepackage{algorithmic}
\usepackage{color}
\usepackage{url}
\usepackage{scrtime}
\usepackage{array}
\usepackage{footnote}
\makesavenoteenv{tabular}
\makesavenoteenv{table}

\author{\authorblockN{ Derrick Wing Kwan Ng,~\IEEEmembership{Member,~IEEE,}
and Robert Schober,~\IEEEmembership{Fellow,~IEEE}\thanks{ This paper has been  presented  in part at  IEEE Globecom 2014 \cite{CN:Kwan_globecom2014} and ICC 2014 \cite{CN:kwan_vicky}, respectively. Derrick Wing Kwan Ng and Robert Schober are with the Institute for Digital Communications (IDC),
Friedrich-Alexander-University Erlangen-N\"urnberg (FAU), Germany (email:\{wingn, rschober\}@ece.ubc.ca). The authors are also with the University of British Columbia, Vancouver, Canada. Derrick Wing Kwan Ng was supported by the Qatar National Research Fund (QNRF), under project NPRP 5-401-2-161. Robert Schober was supported by the AvH Professorship Program of the Alexander von Humboldt Foundation.}}\vspace*{-3mm}}

\title{Secure and Green SWIPT in Distributed Antenna Networks with Limited Backhaul Capacity}

\newtheorem{Thm}{Theorem}

\newtheorem{proposition}{Proposition}
\newtheorem{Remark}{Remark}
 \newcommand{\qed}{\hfill \ensuremath{\blacksquare}}
 
\DeclareMathOperator{\Tr}{\mathrm{Tr}}
\DeclareMathOperator{\zero}{\mathbf{0}}
\DeclareMathOperator{\Rank}{\mathrm{Rank}}

\DeclareMathOperator{\diag}{\mathrm{diag}}
\DeclareMathOperator{\vect}{\mathrm{vec}}
\DeclareMathOperator{\maxo}{maximize}
\DeclareMathOperator{\mino}{minimize}
\DeclareMathOperator{\bigo}{\cal O}

\newcommand{\abs}[1]{\lvert#1\rvert}
\newcommand{\norm}[1]{\lVert#1\rVert}
\newcommand{\bnorm}[1]{\Big\lVert#1\Big\rVert}
\newcolumntype{L}{>{\arraybackslash\raggedright}m{10cm}}

\begin{document}

\maketitle

\begin{abstract}
This paper studies the resource allocation algorithm design for secure information and renewable green energy transfer  to mobile receivers in  distributed antenna  communication systems.  In particular, distributed remote radio heads (RRHs/antennas) are connected to a central processor (CP) via capacity-limited backhaul links to facilitate joint transmission.
The RRHs and the CP are equipped with renewable energy harvesters and  share their energies via a lossy micropower grid  for improving the efficiency in conveying information and green energy to mobile receivers via radio frequency (RF) signals. The considered resource allocation algorithm design is formulated as a mixed non-convex and combinatorial optimization problem taking into account the limited backhaul capacity and the quality of service requirements for simultaneous wireless information and power transfer (SWIPT). We aim at minimizing the total network transmit power when only imperfect channel state information of the wireless  energy harvesting receivers, which have to be powered by the wireless network, is available at the CP. In light of the intractability of the problem, we reformulate it as an optimization problem with binary selection, which facilitates the design of an
iterative resource allocation algorithm to solve the problem optimally using the generalized Bender's decomposition (GBD). Furthermore, a suboptimal algorithm is proposed to strike a balance between computational complexity and system performance.  Simulation results illustrate that the proposed GBD based algorithm obtains the global optimal solution and the suboptimal algorithm achieves a close-to-optimal performance. Besides, the distributed antenna network for SWIPT with renewable energy sharing  is shown to require a lower transmit power compared to a traditional system with  multiple co-located antennas.
\end{abstract}


\section{Introduction} \label{sect1}
 \IEEEPARstart{N}{ext} generation wireless communication systems are required to provide high speed, high  security,  and ubiquitous  communication  with guaranteed quality of service (QoS). These requirements have led to a tremendous energy consumption in both transmitters and receivers. Multiple-input multiple-output (MIMO) technology has emerged as a viable solution for reducing the system
power consumption. In particular,  multiuser MIMO, where a transmitter equipped with multiple antennas serves multiple single-antenna receivers,  is considered  to be an effective  solution for realizing the performance gain offered by multiple antennas. On the other hand, energy harvesting based mobile communication system design  facilitates  self-sustainability for energy
limited communication networks. For instance, the integration of energy harvesting devices into base stations for scavenging energy from traditional renewable energy sources such as  solar and  wind  has been proposed for providing green communication services \cite{Web:Green}--\nocite{JR:Kwan_hybrid_BS}\cite{JR:CoMP_energy}.  However,   theses  natural energy  sources are usually location and climate dependent and may not be suitable for portable mobile receivers.

Recently, wireless power transfer  has been proposed as an emerging alternative energy source,
 where the receivers scavenge energy from the ambient radio frequency (RF)  signals \cite{Ding2014}--\nocite{JR:SWIPT_mag,Krikidis2014,CN:Shannon_meets_tesla,JR:MIMO_WIPT,
 JR:WIPT_fullpaper,JR:EE_SWIPT_Massive_MIMO,JR:energy_beamforming,JR:Kwan_secure_imperfect,JR:rui_zhang,CN:PHY_SEC_max_min}\cite{JR:MOOP}. The broadcast nature of wireless channels  facilitates one-to-many wireless charging, which eliminates the need for power cords and manual recharging, and enables the possibility  of simultaneous wireless information and power transfer (SWIPT). The introduction of an RF energy harvesting capability at the receivers  leads to many interesting and challenging new research problems which have to be solved  to bridge the gap between theory and practice. In \cite{CN:Shannon_meets_tesla} and \cite{JR:MIMO_WIPT}, the fundamental trade-off between  harvested energy and wireless channel capacity was studied for point-to-point and    multiple-antenna wireless broadcast systems, respectively. In   \cite{JR:WIPT_fullpaper},    it was shown that RF energy harvesting can improve  the  energy efficiency of communication networks.  In \cite{JR:EE_SWIPT_Massive_MIMO}, the authors solved the energy efficiency maximization problem for large scale multiple-antenna SWIPT systems. In \cite{JR:energy_beamforming}, the optimal energy transfer dowlink duration was optimized to  maximize the uplink average information transmission rate.   The combination of physical (PHY) layer security and SWIPT was recently investigated in \cite{JR:Kwan_secure_imperfect}--\nocite{JR:rui_zhang,CN:PHY_SEC_max_min}\cite{JR:MOOP} for total transmit power minimization, secrecy rate maximization,  max-min fair optimization, and multi-objective optimization, respectively. {Nevertheless, despite the promising results in the literature
\cite{CN:Shannon_meets_tesla}--\cite{JR:MOOP}, the performance of wireless power/energy transfer systems is still severely limited by the
distance between the transmitter(s) and the receiver(s) due to the high signal attenuation caused by
path loss and shadowing, especially in outdoor environments. Thus,  an exceedingly  large  transmit power is required  to provide QoS in information and power transfer. Hence,  the energy consumption at the transmitters of wireless power transfer systems will become a financial burden to service providers if the efficiency
of wireless power transfer cannot be improved and the energy cost at the transmitters cannot be
reduced.}

In this context, distributed antennas  are an attractive technique for  reducing network power consumption and extending service coverage    \cite{JR:comp}\nocite{JR:comp2,JR:limited_backhaul,JR:Quek}--\cite{CN:Wei_yu_sparse_BF}. { A promising option for the system architecture of  distributed antenna  networks  is the splitting of the functionalities of the base station between a central processor (CP)  and a set of low-cost remote
radio heads (RRHs). In particular, the  CP performs the power hungry and computationally intensive baseband signal processing while the RRHs are responsible for all RF operations such as analog filtering and power amplification. The RRHs are distributed across the network and connected to the CP via backhaul links.   This system architecture is known as ``Cloud Radio Access Network" (C-RAN) \cite{JR:Cloud,JR:Cloud2,JR:Cloud3}.} The distributed antenna  system architecture reduces the distance between transmitters and receivers.
Furthermore, it inherently provides  spatial diversity   for combating  path loss and shadowing. It has been shown  in  \cite{JR:comp,JR:comp2} that  distributed antenna  systems with full cooperation between the transmitters achieve  a superior performance compared to co-located antenna systems. Yet, transferring the information data of all users from the CP to all RRHs, as is required for full cooperation,  may be infeasible when the capacity of the backhaul links is limited. Hence, resource allocation for distributed antenna  networks  with finite backhaul capacity has attracted considerable
attention in the research community \cite{JR:limited_backhaul}--\cite{CN:Wei_yu_sparse_BF}. In \cite{JR:limited_backhaul}, the authors
studied the energy efficiency of distributed antenna  multicell networks with capacity constrained backhaul links.
In \cite{JR:Quek} and \cite{CN:Wei_yu_sparse_BF}, iterative algorithms were proposed to reduce the total system backhaul capacity
 consumption while guaranteeing  reliable communication to the mobile users. However, the problem formulations in \cite{JR:Quek} and \cite{CN:Wei_yu_sparse_BF} do not constrain the capacity consumption of individual backhaul links which may lead to an information overflow in some of the backhaul links. Moreover,   \cite{JR:limited_backhaul}--\cite{CN:Wei_yu_sparse_BF}  assume the availability of an  ideal power supply for each RRH such that a large amount of energy can be continuously used for operation of the system whenever needed. However, assuming availability of an ideal power supply for the RRHs may not be realistic in practice, especially in developing countries or remote areas \cite{JR:limited_backhaul}--\cite{CN:Wei_yu_sparse_BF}. In addition,  the receivers
in \cite{JR:comp}\nocite{JR:comp2,JR:limited_backhaul,JR:Quek}--\cite{CN:Wei_yu_sparse_BF} were assumed to be powered by constant energy sources which may also not be a valid assumption for energy-limited handheld
devices. Although the transmitters can be powered by renewable green energy and the  signals transmitted in the RF by the RRHs could be exploited as energy sources to the  receivers for extending
their  lifetimes, resource allocation algorithm design for  utilizing green energy in distributed antenna SWIPT systems has not been considered in the literature, yet.

Motivated by the aforementioned observations, in this paper, we propose the use of distributed antenna  communication networks for transferring information and  green  renewable energy to mobile receivers wirelessly.  We formulate the resource allocation algorithm design as a non-convex optimization problem. Taking into account the limited backhaul capacity,  the harvested renewable energy sharing between RRHs, and the imperfect CSI of the energy harvesting receivers, we   minimize  the total network transmit power while ensuring the QoS of the wireless receivers for both secure communication and efficient wireless power transfer.  To this end, we propose an optimal iterative algorithm based on the generalized Bender's decomposition. In addition, we propose a low complexity suboptimal resource allocation scheme based on the difference of convex functions (d.c.) programming  which provides a locally optimal solution for the considered optimization problem.

\section{System Model}
\label{sect:system model}

\subsection{Notation}
We use boldface capital and lower case letters to denote matrices and vectors, respectively. $\mathbf{A}^H$, $\Tr(\mathbf{A})$, and $\Rank(\mathbf{A})$ represent the  Hermitian transpose, the trace, and the rank of  matrix $\mathbf{A}$, respectively; $\mathbf{A}\succ \mathbf{0}$ and $\mathbf{A}\succeq \mathbf{0}$ indicate that $\mathbf{A}$ is a positive definite and a  positive semidefinite matrix, respectively; $\vect(\mathbf{A})$ denotes the vectorization of matrix $\mathbf{A}$ by stacking its columns from left to right to form a column vector. $\mathbf{I}_N$ is the $N\times N$ identity matrix; $\mathbb{C}^{N\times M}$ and $\mathbb{R}^{N\times M}$ denote the set of all $N\times M$ matrices with complex and real entries, respectively; $\mathbb{H}^N$ denotes the set of all $N\times N$ Hermitian matrices; $\diag(x_1, \cdots, x_K)$ denotes a diagonal matrix with the diagonal elements given by $\{x_1, \cdots, x_K\}$; $\abs{\cdot}$ and $\norm{\cdot}_p$ denote the absolute value of a complex scalar and the
$l_p$-norm of a vector, respectively. In particular, $\norm{\cdot}_0$ is known as the $l_0$-norm of a vector and denotes the number of non-zero entries in the vector; the circularly symmetric complex Gaussian (CSCG) distribution is denoted by ${\cal CN}(\mu,\sigma^2)$ with mean  $\mu$ and variance $\sigma^2$; $\sim$ stands for ``distributed as";  $\big[x\big]^+=\max\{0,x\}$; $\mathbf{1}$ denotes a column vector with all elements equal to one. $\big[\cdot\big]_{a,b}$ returns the $(a,b)$-th element of the input matrix, $\boldsymbol{\theta}_n$ is the
   $n$-th unit column vector, i.e., $\big[\boldsymbol{\theta}_n\big]_{t,1}=1, t= n,$ and $\big[\boldsymbol{\theta}_n\big]_{t,1}=0,\forall t\ne n$; and for a real valued  continuous function $f(\cdot)$,
 $\nabla_{\mathbf{x}} f(\mathbf{x})$ represents the gradient of $f(\cdot)$ with respect to vector $\mathbf{x}$.

\begin{figure*}
\centering
\includegraphics[scale=0.6]{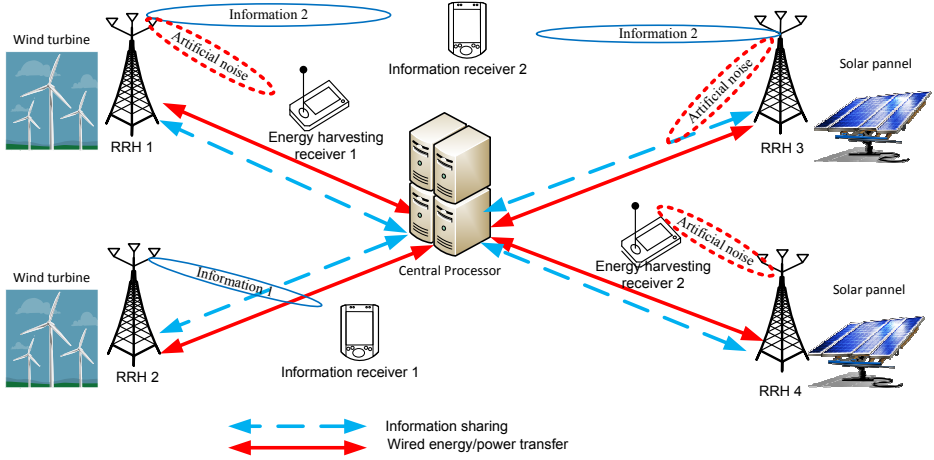}
\caption{Distributed antenna  multiuser downlink communication system model with a central processor (CP), $L=4$ remote radio heads (RRHs), $K=2$  information receivers (IRs), and $M=2$ energy harvesting receivers (ERs). The blue solid ellipsoids represent the information signals intended for the different IRs.  The red dotted ellipsoids illustrate the
dual functionality of artificial noise in providing security and facilitating efficient energy transfer to the ERs.}
\label{fig:system_model}
\end{figure*}
\subsection{Distributed Antenna  System Model and Central Processor}
\label{sect:multicell-central-unit}
We consider a distributed antenna  multiuser downlink communication network. The system consists of a CP,  $L$ RRHs,   $K$ information receivers (IRs), and $M$ energy harvesting receivers (ERs), cf. Figure \ref{fig:system_model}. Each RRH is equipped with $N_\mathrm{T}>1$ transmit antennas. The IRs and ERs are single antenna devices  which exploit the received signal powers in the RF for information decoding and energy harvesting, respectively. In practice, the ERs may be idle IRs which are scavenging  energy from the RF to extend their lifetimes. On the other hand, the CP is the core unit of the network, which has the data intended for all IRs. Besides, we assume that all computations are performed in the CP. In particular, based on the available CSI, the CP computes the
resource allocation policy and broadcasts it to  all RRHs. Each RRH receives the control signals for resource allocation and the data of the $K$ IRs from the CP via a backhaul link. The backhaul links can be implemented with different last-mile communication technologies such as digital subscriber line (DSL) or out-of-band microwave links. Thus, the backhaul capacity may be limited.
Furthermore, we assume that the CP is integrated with a constant energy source (e.g., a diesel generator) for supporting its normal operation, and the distributed RRHs are equipped with traditional energy harvesters such as solar panels and wind  turbines for generation of renewable energy.
The harvested energy  can be exchanged between the CP and the RRHs over a micropower grid and
the  CP  manages the energy flow in the micropower grid\footnote{The proposed system can be viewed as a hybrid information and energy distribution network. In particular, the  green energy harvested at the RRHs is shared via the micro-grid and distributed to the ERs via RF.    }, cf. Section \ref{sect:power_supply}.

\subsection{Channel Model}

We focus on a frequency flat fading channel and a time division duplexing (TDD) system.  The wireless information and power transfer from the RRHs to the receivers is divided into time slots.
 The received signals at IR $k\in\{1,\ldots,K\}$ and ER $m\in\{1,\ldots,M\}$ in one time slot are given by
\begin{eqnarray}
y_{k}^{\mathrm{IR}}=\mathbf{h}_k^H\mathbf{x}+ n^{\mathrm{IR}}_{k}\quad \mbox{and}\quad y_{m}^{\mathrm{ER}}=\mathbf{g}_m^H\mathbf{x}+n^{\mathrm{ER}}_{m},\,\,
\end{eqnarray}
respectively, where $\mathbf{x}\in\mathbb{C}^{N_{\mathrm{T}}L\times1}$ denotes the joint transmit vector of the $L$ RRHs to the $K$ IRs and the $M$ ERs. The channel between the $L$ RRHs  and  IR $k$ is denoted by $\mathbf{h}_k\in\mathbb{C}^{N_{\mathrm{T}}L\times1}$, and we use $\mathbf{g}_m\in\mathbb{C}^{N_{\mathrm{T}}L\times1}$ to denote the channel  between the $L$ RRHs  and   ER  $m$. We note that the channel vector captures the joint effects of multipath fading and path loss.  { $n^{\mathrm{IR}}_{k}$ and  $n^{\mathrm{ER}}_{m}$ include the joint effects  of thermal noise, signal processing
noise, and possibly present received multicell interference  at IR $k$ and ER $m$, respectively, and are  modeled as additive white Gaussian noise (AWGN) with zero mean and variances $\sigma_{\mathrm{IR}_k}^2$ and $\sigma_{\mathrm{ER}_m}^2$, respectively. }

\subsection{Channel State Information}
We assume that
 $\mathbf{h}_k,\forall k\in\{1,\ldots,K\}$, and $\mathbf{g}_{m},\forall m\in\{1,\ldots,M\}$, can be reliably obtained  at the beginning of each scheduling slot by exploiting the  channel reciprocity and the pilot sequences in the handshaking signals exchanged between the RRHs and the receivers. Besides,  the estimate of $\mathbf{h}_k$ is refined at the CP  during the entire scheduling slot  based on the pilot sequences contained in acknowledgement  packets. As a result,  we can assume that the CSI  for the RRHs-to-desired IR  links is perfect  during the entire transmission period. In contrast, the ERs do not interact with the RRHs during information transmission.  Thus, the CSI of the ERs may be outdated during transmission and we use  a deterministic model \cite{JR:Kwan_secure_imperfect,JR:Robust_error_models1} for characterizing the resulting CSI uncertainty. More precisely, the  CSI of the link between the RRHs
and ER $m$ is given by
\begin{eqnarray}\label{eqn:outdated_CSI}
\mathbf{g}_m&=&\mathbf{\hat g}_m + \Delta\mathbf{g}_m,\,   m\in\{1,\ldots,M\}, \,\mbox{and}\notag\\
{\Omega }_m&\triangleq& \Big\{\Delta\mathbf{g}_m\in \mathbb{C}^{N_{\mathrm{T}}L\times 1}  :\Delta\mathbf{g}_m^H\mathbf{\Xi}_m  \Delta\mathbf{g}_m \le \varepsilon_m^2\Big\},\label{eqn:outdated_CSI-set}
\end{eqnarray}
where $\mathbf{\hat g}_m\in\mathbb{C}^{N_{\mathrm{T}}L\times 1}$ is the channel estimate of ER $m$ available at the CP at the beginning of a scheduling slot. $ \Delta\mathbf{g}_m$ represents the unknown channel uncertainty  of ER $m$ due to the slowly time varying nature of the channel during transmission. In (\ref{eqn:outdated_CSI-set}), we define  set ${\Omega }_m$  which contains all possible CSI uncertainties of ER $m$. Specifically, ${\Omega }_m$ specifies an ellipsoidal uncertainty region for the estimated CSI of ER  $m$, where $\varepsilon_m>0$ and  $\mathbf{\Xi}_m\in\mathbb{C}^{N_{\mathrm{T}}L\times N_{\mathrm{T}}L},\mathbf{\Xi}_m\succ \zero$ represent the radius  and  the orientation
of the region, respectively. For instance, (\ref{eqn:outdated_CSI-set}) represents an Euclidean sphere when $\mathbf{\Xi}_m=\mathbf{I}_{N_{\mathrm{T}}L}$.
 In practice,  the value of $\varepsilon_m^2$ depends on the coherence time of the associated channel and $\mathbf{\Xi}_m$ depends on the adopted channel estimation method.
\begin{figure*}[t]
\centering
\includegraphics[scale=0.8]{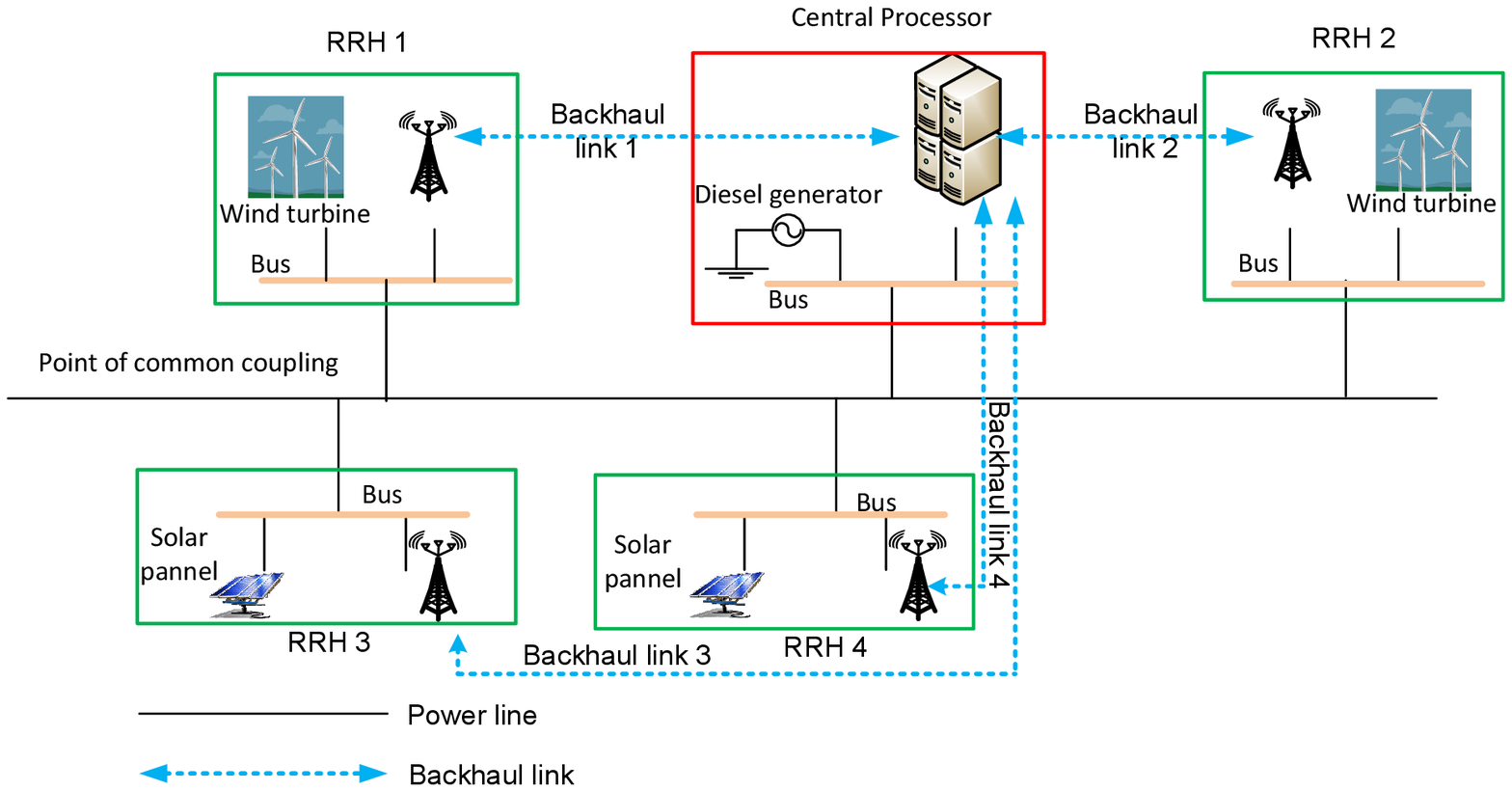}
\caption{A simplified micropower grid model with a \emph{point of common coupling} connecting a  central processor (CP) and $L=4$ remote radio heads (RRHs). The black solid and blue dashed lines indicate the power line and backhaul connections, respectively.}\vspace*{-4mm}
\label{fig:grid_model}
\end{figure*}

\subsection{Signal and Backhaul Models}
In each scheduling time slot, $K$ independent signal streams are transmitted simultaneously to the $K$ IRs. Specifically,  a dedicated beamforming vector, $\mathbf{w}_k^l\in\mathbb{C}^{N_{\mathrm{T}}\times1}$, is allocated to IR $k$ at RRH $l\in\{1,\ldots,L\}$ to facilitate information transmission. For  the sake of presentation, we define a  super-vector $\mathbf{w}_k\in\mathbb{C}^{N_{\mathrm{T}}L\times 1}$  for IR $k$ as
\begin{eqnarray}
\mathbf{w}_k=\vect\big([\mathbf{w}_k^1 \,\mathbf{w}_k^2\,\ldots\,\mathbf{w}_k^L]\big).
\end{eqnarray}
 Here, $\mathbf{w}_k$ represents the joint beamformer used by the $L$ RRHs for serving IR $k$. Then, the information signal to IR $k$,  $\mathbf{x}_k$, can be expressed as
\begin{eqnarray}
\mathbf{x}_k=\mathbf{w}_k d_k,
\end{eqnarray}
where $d_k\in\mathbb{C}$ is the data symbol for IR $k$ and  ${\cal E}\{\abs{d_k}^2\}=1,\forall k\in\{1,\ldots,K\}$, is assumed without loss of generality. { The information signals intended for the desired IRs can be overheard by the ERs that are in the
range of service coverage. Since the  ERs may be malicious, they may
eavesdrop the information signal of the selected IRs. This has to be taken into account
for resource allocation design for providing secure communication services in the considered distributed antenna network. Thus, to guarantee communication security, the RRHs have
to employ a resource allocation algorithm that accounts for this unfavourable scenario
and treat the ERs as potential eavesdroppers\footnote{Although the ERs are low-power devices,
malicious ERs do not have to decode the eavesdropped information in
real time. They can act as information collectors which sample the received signals and
store them for future decoding by other energy unlimited and computationally powerful
devices.}, see also \cite{JR:Kwan_secure_imperfect,JR:rui_zhang,CN:PHY_SWIPT1,CN:ken_PHY_SWIPT}.} To this end,  artificial noise  is transmitted by the RRHs\footnote{{
In \cite{JR:phy_channel_inversion}, the  secrecy rate achievable with regularized channel inversion for large numbers of users and transmit antennas was studied. However, the method proposed in \cite{JR:phy_channel_inversion} can  guarantee a strictly positive secrecy rate only if the number of transmit antennas tends to infinity.}} which can be used to degrade the channels between
the RRHs and the potential eavesdroppers and to serve as an energy source for the ERs. {
Hence, the transmit signal vector $\mathbf{x}$ at the RRHs is given by
\begin{eqnarray}
\mathbf{x}=\underbrace{ \sum_{k=1}^K\mathbf{x}_k}_{\mbox{desired information signal}}+\underbrace{\mathbf{v}}_{\mbox{artificial noise}},
\end{eqnarray}
where $\mathbf{v}\in\mathbb{C}^{N_{\mathrm{T}}L\times 1}$ is the artificial noise vector generated by the RRHs and  modeled as a complex Gaussian random vector, i.e.,
$
\mathbf{v}\sim {\cal CN}(\mathbf{0}, \mathbf{V})$, where $\mathbf{V}\in \mathbb{H}^{N_{\mathrm{T}}L}, \mathbf{V}\succeq \mathbf{0}$, denotes the covariance matrix of $\mathbf{v}$.  The artificial noise  $\mathbf{v}$ interferes the IRs and ERs since $\mathbf{v}$ is  unknown to both types of receivers. Hence,  artificial noise transmission has to be carefully designed to degrade the channels of the ERs  while  having a minimal effect on the IRs.}  {
In fact, the covariance matrix of the artificial noise will be optimized under the proposed optimization framework.  We note that  artificial noise vector $\mathbf{v}$ can be generated locally at the RRHs and does not have  to be sent via the backhaul links.
} On the other hand, the data of each IR is delivered from the CP to the RRHs via backhaul links. The backhaul capacity consumption for backhaul link $l\in\{1,\ldots,L\}$ is given by
\begin{eqnarray}{
C^{\mathrm{Backhaul}}_l=\sum_{k=1}^K\bnorm{\norm{\mathbf{w}_k^l}_2}_0\,\, R_{\mathrm{B}_k},}
\end{eqnarray}
where $R_{\mathrm{B}_k}$ is the required backhaul data rate for conveying the data of IR $k$ to a RRH and $\sum_{k=1}^K\bnorm{\norm{\mathbf{w}_k^l}_2}_0$ counts the number of IRs consuming the capacity of backhaul $l$. We note that the backhaul links  may be capacity-constrained and the CP may not be able to send the data of all IRs to all RRHs as required for
full cooperation. Thus, to reduce the load on the backhaul links, the CP can enable partial cooperation by sending the data of IR $k$  only to a subset of  the RRHs. In particular, by setting $\mathbf{w}_l^k=\mathbf{0}$, RRH $l$ is not participating in the joint data transmission to IR $k$.  Thus, the CP is not required to send the data for IR $k$ to  RRH $l$  via the backhaul link which leads to a lower information flow in the backhaul link.

\subsection{RRH Power Supply Model}
\label{sect:power_supply}
The constant energy source of the CP transfers energy to all RRHs via a dedicated power grid (micropower grid)  for supporting the power consumption at the RRHs and
facilitating a  more efficient network operation, cf. Figure \ref{fig:grid_model}. In particular,  a bus in Figure \ref{fig:grid_model} refers to the internal power line connection of zero impedance between two elements. The CP is connected  to a \emph{point of common coupling}
to convey  energy to the micropower grid and has full control over the micropower grid.

Since each RRH is equipped with  energy harvesters for harvesting renewable energy,    the energy harvested by  the RRHs can also be shared in the communication system via the micropower grid. By exploiting the spatial diversity inherent to the distributed antenna network also for energy harvesting,  we can overcome potential  energy harvesting imbalances in the network for improving the system performance.  In other words, there are $L+1$ energy sources for supporting the CP and the $L$ RRHs. We denote the unit of energy  transferred from energy source $n\in\{1,\ldots,L+1\}$ to the micro-grid as $E^{\mathrm{S}}_n$ where the power generator at the CP is the $(L+1)$-th energy source.  The power loss in delivering the power from all the $L+1$ energy sources to the $L$ RRH   is given by  \cite{book:B_matrix}
  \begin{eqnarray}\label{eqn:B_matrix}
  P_{\mathrm{Loss}}=\sum_{n=1}^{L+1}\sum_{m=1}^{L+1}  E^{\mathrm{S}}_l B_{n,m} E^{\mathrm{S}}_m= (\mathbf{e}^{\mathrm{S}} )^T \mathbf{B}\mathbf{e}^{\mathrm{S}}>0,
   \end{eqnarray}
   where $\mathbf{e}^{\mathrm{S}} =[E^{\mathrm{S}}_1\, E^{\mathrm{S}}_2\,\,\ldots\,E^{\mathrm{S}}_{L+1}]^T$, $\mathbf{e}^{\mathrm{S}}\in\mathbb{R}^{L+1}$. $B_{n,m}=[\mathbf{B}]_{n,m}$  is known as the B-coefficient and
    $ \mathbf{B}\in\mathbb{R}^{(L+1)\times (L+1)}$, $\mathbf{B}\succ\zero$, is the
   \emph{B-coefficient matrix}  \cite{book:B_matrix} which takes into account the distance dependent power line resistance, the phase angles of the electrical currents, and the voltages generated by the different energy sources. We note that the  \emph{B-coefficient matrix} is a constant for a fixed number of loads and a fixed grid connection topology. We assume that the  \emph{B-coefficient matrix} is known to the CP for energy allocation from long term measurements. Furthermore, the  energy supplied by energy source $n$ is given by
  \begin{eqnarray}\label{eqn:maximum_energy_supply}
  \mbox{Supplied energy: } \underbrace{\boldsymbol{\theta}_n^T\mathbf{e}^{\mathrm{S}}}_{E^{\mathrm{S}}_n}\le E^{\max}_n, \forall n\in\{1,\ldots,L+1\},
 \end{eqnarray}
   where $E^{\max}_n$  is the maximum  energy available  at energy source  $n$ and represents the total amount of energy
   generated by energy source $n$.    In this paper, each energy source is able to adjust the amount of energy injected into the micropower grid.

    {
In practice, the coherence time of the communication  channel is much shorter than that of the renewable energy harvesting process at the RRHs.
 For instance, for a carrier center  frequency  of $915$ MHz and $1.4$ m/s receiver speed, the coherence time for wireless communication is in the order of $100$ ms.  In other words, the resource allocation policy has to be updated roughly every $100$ ms.  On the other hand, the renewable energy arrival rate at the energy harvesters of the RRHs changes relatively slowly. For example,  solar energy and wind energy change in the order of a few tens of seconds  \cite{JR:CoMP_energy}. Thus,  for resource allocation design, we assume that  $E^{\max}_n$ in (\ref{eqn:maximum_energy_supply}) is a known constant}. {
  Furthermore,  we focus on the resource allocation for small cell systems, i.e., the inter-site distances  between the RRHs is in the order of hundreds of meters. Thus, the energy propagation delay between two renewable energy harvesters  is less than $1\mu$s, which is  negligibly  small compared to the coherence time of the communication channel and can be neglected in the power supply model.}

\section{Problem Formulation}
\label{sect:forumlation}
In this section, we define  the QoS
metrics for the design of secure communication and power efficient wireless energy transfer. Then, the resource allocation algorithm design is formulated as a non-convex optimization problem.

\subsection{Achievable Data Rate and RF Energy Harvesting}
\label{subsect:Instaneous_Mutual_infxormation}
The  achievable data rate (bit/s/Hz) between the $L$ RRHs and IR $k$ is given by
\begin{eqnarray}
C_{k}&=&\log_2(1+\Gamma_{k}),\,\,
\mbox{where}\quad \\
\Gamma_{k}&=&\frac{\abs{\mathbf{h}_k^H\mathbf{w}_k}^2}{\sum\limits_
{\substack{j\neq k}}^K\abs{\mathbf{h}_k^H\mathbf{w}_j}^2+\Tr(\mathbf{V}\mathbf{h}_k \mathbf{h}_k^H)+\sigma_{\mathrm{IR}_k}^2}
\end{eqnarray}
is the receive signal-to-interference-plus-noise ratio (SINR) at IR $k$.

{ Since the computational capability of the ERs (potential eavesdroppers) is not known at the CP, we consider the worst-case scenario for providing communication security. Specifically, in the worst case, the ERs  are able to
remove all multiuser interference and multicell interference via successive interference cancellation  before attempting to decode the information of desired IR $k$.} Therefore,  the achievable data rate between the RRHs and ER (potential eavesdropper) $m$  is given  by
\begin{eqnarray}\label{eqn:cap-eavesdropper}
C_{\mathrm{ER}_m}&=&\log_2\Big(1+\Gamma_{\mathrm{ER}_m}\Big)\,\,\,\,
\mbox{and}\,\,\\ \label{eqn:SINR_up_idle}
\Gamma_{\mathrm{ER}_m}&=&\frac{\abs{\mathbf{g}_m^H\mathbf{w}_k}^2 }{\sum_{j\ne k}\abs{\mathbf{g}_m^H\mathbf{w}_j}^2 +\Tr(\mathbf{V}\mathbf{g}_m\mathbf{g}_m^H)+\sigma_{\mathrm{ER}_m}^2}\notag\\
&  \stackrel{(a)}{\le}& \frac{\abs{\mathbf{g}_m^H\mathbf{w}_k}^2 }{\Tr(\mathbf{V}\mathbf{g}_m\mathbf{g}_m^H)+\sigma_{\mathrm{s}}^2 }   ,
   \end{eqnarray}
where $\Gamma_{\mathrm{ER}_m}$ is  the received SINR at ER $m$ and $\sigma_{\mathrm{s}}^2$ is the joint  power of the  signal processing noise and the thermal noise. $(a)$ reflects the aforementioned worst-case assumption\footnote{We note that the proposed framework can be easily extended to the case when a single-user detector is employed at the potential eavesdroppers. This modification does not change the structure of the problem and does not affect the resource allocation algorithm design.} and constitutes an upper bound on the received SINR at ER $m$ for decoding the information of IR $k$.

In the considered system, the information signal, $\mathbf{w}_k d_k,\forall k\in\{1,\ldots,K\}$,  serves as a dual purpose carrier for both information and energy. Besides, the artificial noise signal also acts as an energy source to the ERs. The total amount of energy\footnote{We adopt the normalized energy unit  Joule-per-second in this paper.  Therefore,
the terms ``power" and ``energy" are used interchangeably.} harvested by ER $m\in\{1,\ldots,M\}$ is given by
\begin{eqnarray}\label{eqn:ER_power}
E_{m}^{\mathrm{ER}}=\mu\Big(\Tr(\mathbf{V}\mathbf{g}_m\mathbf{g}_m^H)+\sum_{k=1}^K\abs{\mathbf{g}_m^H\mathbf{w}_k}^2\Big),
\end{eqnarray}
where $0<\mu\leq1$ denotes the efficiency of converting the received RF energy to electrical energy for storage. We assume that $\mu$ is a constant and is identical for all ERs. We note that the contribution of the antenna thermal noise power and the multicell interference power to the harvested energy  is negligibly small compared to the energy harvested  from the received signal, $\Tr(\mathbf{V}\mathbf{g}_m\mathbf{g}_m^H)+\sum_{k=1}^K\abs{\mathbf{g}_m^H\mathbf{w}_k}^2$, and thus is neglected in (\ref{eqn:ER_power}).
\subsection{Optimization Problem Formulation}
\label{sect:cross-Layer_formulation}
The system objective is to minimize  the total network transmit power  while providing QoS for reliable communication and efficient power transfer in a given time slot for given maximum backhaul capacities. The resource allocation algorithm design  is formulated as the following optimization problem\footnote{{ For resource allocation algorithm design,  we assume that the  problem in (\ref{eqn:cross-layer}) is feasible. In practice, the probability that (\ref{eqn:cross-layer}) is feasible can be improved by a suitable scheduling of the IRs and ERs in the media access control layer. }}:
\begin{eqnarray} \label{eqn:cross-layer}\notag
&&\hspace*{10mm} \underset{ \mathbf{V}\in\mathbb{H}^{N_{\mathrm{T}}L},\mathbf{e}^{\mathrm{S}},\mathbf{w}_k}{\mino}\,\, \sum_{k=1}^K\sum_{l=1}^L\norm{\mathbf{w}^l_k}_2^2 +\Tr(\mathbf{V})\\
\notag \mathrm{s.t.}\hspace*{-2mm} &&\hspace*{-2mm}\mbox{C1: }\frac{\abs{\mathbf{h}_k^H\mathbf{w}_k}^2}{\sum\limits_
{\substack{j\neq k}}^K\abs{\mathbf{h}_k^H\mathbf{w}_j}^2+\Tr(\mathbf{V}\mathbf{h}_k\mathbf{h}_k^H)+\sigma_{\mathrm{IR}_k}^2}\ge\Gamma_{\mathrm{req}_k},\,\, \forall k, \notag\\
&&\hspace*{-2mm}\mbox{C2: }{\max_{\Delta\mathbf{g}_m\in{\Omega }_m}} \frac{\abs{\mathbf{g}_m^H\mathbf{w}_k}^2 }{\Tr(\mathbf{V}\mathbf{g}_m\mathbf{g}_m^H)+\sigma_{\mathrm{s}}^2 }  \le\Gamma_{\mathrm{tol}},\, \forall m, k,\notag\\
&&\hspace*{-2mm}\mbox{C3: } { \sum_{k=1}^K\bnorm{\norm{\mathbf{w}_k^l}_2}_0\, R_{\mathrm{B}_k}\le C_{l}^{\mathrm{B}_{\max}},\,\, \forall l,\notag}\\
&&\hspace*{-2mm}\mbox{C4: } P_{\mathrm{C}}^\mathrm{CP}+\sum_{l=1}^L \Big\{P_{\mathrm{C}_l}+ \rho\Big(\sum_{k=1}^K\norm{\mathbf{w}^l_k}^2_2+\Tr(\mathbf{V}\mathbf{R}_l) \Big)\Big\}\notag\\
&&\le \mathbf{1}^T\mathbf{e}^{\mathrm{S}}- (\mathbf{e}^{\mathrm{S}} )^T\mathbf{B}\mathbf{e}^{\mathrm{S}},\notag\\
&&\hspace*{-2mm}\mbox{C5: }\boldsymbol{\theta}_n^T\mathbf{e}^{\mathrm{S}}\le E^{\max}_n,\,\, \forall n\in\{1,\ldots,L+1\},\notag\\
&&\hspace*{-2mm}\mbox{C6: }\Tr(\mathbf{V}\mathbf{R}_l)+\sum_{k=1}^K\norm{\mathbf{w}^l_k}^2_2\le P^{\mathrm{T}_{\max}}_l,\,\, \forall l,\notag\\
&&\hspace*{-2mm}\mbox{C7:}\,\, \min_{\Delta\mathbf{g}_m\in{\Omega }_m} E_{m}^{\mathrm{ER}}\ge P^{\min}_{m},\,\, \forall m,\,\, \,\, \,\,\notag\\
&&\hspace*{-2mm}\mbox{C8:}\,\, \mathbf{e}^{\mathrm{S}}\ge \zero ,\,\, \,\, \,\,\mbox{C9:}\,\, \mathbf{V}\succeq \zero ,
\end{eqnarray}
where $\mathbf{R}_l\triangleq\diag\Big(\underbrace{0,\cdots,0}_{(l-1)N_\mathrm{T}},\underbrace{1,\cdots,1}_{N_\mathrm{T}},
\underbrace{0,\cdots,0}_{(L-l)N_\mathrm{T}}\Big),\forall l\in\{1,\ldots,L\}$,
is a block diagonal matrix. $\Gamma_{\mathrm{req}_k}>0$ in constraint C1 indicates the required minimum  receive SINR at IR $k$ for information decoding.  Constraint C2 is imposed such that for a given CSI uncertainty set $\Omega_m$, the maximum received SINR at ER $m$ is less than the maximum tolerable received SINR $\Gamma_{\mathrm{tol}}$. In practice, the CP sets $\Gamma_{\mathrm{req}_k}\gg \Gamma_{\mathrm{tol}}>0,\forall k\in\{1,\ldots,K\}$, to ensure secure communication. Specifically,  the adopted problem formulation guarantees that the achievable secrecy rate for IR $k$ is $R_{\mathrm{sec}_k}= [\log_2(1+\Gamma_{\mathrm{req}_k})-\log_2(1+\Gamma_{\mathrm{tol}})]^+\ge 0$. We note that although $\Gamma_{\mathrm{req}_k}$ and $\Gamma_{\mathrm{tol}}$ in C1 and C2, respectively,  are
not optimization variables in this paper, a balance between
secrecy capacity and system capacity can be struck
by varying their values.  In fact, when constraint C2 is removed from the optimization problem, PHY layer security is not considered in the system. In other words, the adopted problem formulation is a generalized framework which provides flexibility in  controlling the level of communication security. In C3,  the backhaul capacity consumption for backhaul link $l$ is constrained to be less than the maximum  available capacity of backhaul link $l$, i.e., $C_{l}^{\mathrm{B}_{\max}}$. { The corresponding data rate per backhaul link use for IR $k$ is set to the same as the required secrecy rate, i.e.,  $R_{\mathrm{B}_k}=R_{\mathrm{sec}_k}=[\log_2(1+\Gamma_{\mathrm{req}_k})- \log_2(1+\Gamma_{\mathrm{tol}})]^+$}. The right hand side of C4, $\mathbf{1}^T\mathbf{e}^{\mathrm{S}}-(\mathbf{e}^{\mathrm{S}} )^T \mathbf{B}\mathbf{e}^{\mathrm{S}}$,  denotes the maximum  available power in the power grid taking into account the power loss in the power lines. We note that  $\mathbf{1}^T\mathbf{e}^{\mathrm{S}}- (\mathbf{e}^{\mathrm{S}} )^T\mathbf{B}\mathbf{e}^{\mathrm{S}}\ge 0$ always holds by the law of conservation of energy. The left hand side of C4 accounts for the total power consumption in the network.   In C4,  $ P_{\mathrm{C}}^\mathrm{CP}$ and  $P_{\mathrm{C}_l}$ represent the fixed circuit power consumption in the CP and RRH $l$, respectively; the term $\sum_{k=1}^K\norm{\mathbf{w}^l_k}^2_2+\Tr(\mathbf{V}\mathbf{R}_l)$ denotes the output power of the power amplifier of RRH $l$, and $\rho\ge 1$  is a constant accounting for the power inefficiency of the power amplifier;
C5 is a constraint on the maximum power supply from energy source $n\in\{1,\ldots,L+1\}$. Constant $P^{T_{\max}}_l$ in C6 is the maximum transmit power allowance for RRH $l$, which can be used to limit out-of-cell
interference.  $P^{\min}_m$  in C7 is the minimum required power transfer to ER $m$. We note that for given CSI uncertainty sets $\Omega_m,\forall m$, the CP can guarantee the minimum required power transfer to the $M$ ERs only if they use all their received power for energy harvesting.
 C8 is the non-negativity constraint on the energy supply optimization variables.  C9 and $\mathbf{V}\in \mathbb{H}^{N_{\mathrm{T}}L}$ constrain matrix $\mathbf{V}$  to be a positive semidefinite Hermitian matrix, i.e., they ensure that $\mathbf{V}$ is a valid covariance matrix.

\begin{Remark} We  emphasize that the problem formulation considered in this paper is different from that in \cite{JR:Quek} and \cite{CN:Wei_yu_sparse_BF}. In particular, we focus on the capacity consumption of  individual backhaul links while \cite{JR:Quek} and \cite{CN:Wei_yu_sparse_BF} studied the total network backhaul capacity consumption. Besides, we  constrain the capacity consumption of the individual backhaul links which is not possible with the problem formulation adopted in  \cite{JR:Quek} and  \cite{CN:Wei_yu_sparse_BF}. On the other hand, although the combination of PHY layer security and SWIPT has been recently considered in    \cite{JR:Kwan_secure_imperfect} and \cite{JR:rui_zhang}, the results in \cite{JR:Kwan_secure_imperfect} and \cite{JR:rui_zhang}  cannot be directly applied to our problem formulation due to the combinatorial constraints on the limited backhaul capacity and the exchange of harvested power between RRHs\footnote{{ We note that the proposed optimization framework can be extended to include additional passive eavesdroppers, for which instantaneous CSI is not available at the CP, by introducing probabilistic maximum tolerable SINR constraints for the passive eavesdroppers following a similar approach as in \cite{JR:Kwan_secure_imperfect} and  \cite{JR:chance_constraint}.
}} .
\end{Remark}
\begin{Remark}
The proposed framework can be extended to the case of dynamic energy harvesting with energy storage in the RRHs by following similar approaches as in \cite{JR:Kwan_hybrid_BS} and \cite{JR:TCOM_harvesting}. { However, in this paper, we assume that when the renewable energy harvested by the RRHs  exceeds
the total energy consumption of the communication system,  the surplus  harvested renewable energy at the RRHs is transferred  to the external power grid, which is possible  in a smart grid setup \cite{JR:CoMP_energy}.
}
\end{Remark}
\section{Resource Allocation Algorithm Design} \label{sect:RA}
The optimization problem in (\ref{eqn:cross-layer})  is a non-convex  problem.  In the following, we first develop an iterative resource allocation algorithm  for obtaining the global optimal solution based on the generalized Bender's decomposition.  Then, we propose a low computational complexity suboptimal algorithm inspired by the difference of convex functions programming.

\subsection{Problem Reformulation}
In this section, we reformulate the considered optimization problem to facilitate the
 development of resource allocation algorithms. First, we define $\mathbf{W}_k=\mathbf{w}_k\mathbf{w}_k^H$, $\mathbf{H}_k=\mathbf{h}_k\mathbf{h}_k^H$, and $\mathbf{G}_m=\mathbf{g}_m\mathbf{g}_m^H$ for notational simplicity. Besides, we introduce an auxiliary optimization variable $s_{l,k}$ for simplifying the problem. Then, we recast the optimization problem  as follows:
\begin{eqnarray}\label{eqn:equivalent-binary}
&&\hspace*{-0mm} \underset{\mathbf{W}_k,\mathbf{V}\in \mathbb{H}^{N_{\mathrm{T}}L},\mathbf{e}^{\mathrm{S}}, s_{l,k}}{\mino}\,\, \sum_{k=1}^K\Tr(\mathbf{W}_k)+\Tr(\mathbf{V})\notag\\
\mathrm{s.t.} &&\hspace*{-5mm}\mbox{C1: }\hspace*{-0.5mm}\frac{\Tr(\mathbf{H}_k\mathbf{W}_k)}{\Gamma_{\mathrm{req}_k}}\hspace*{-0.5mm}\ge\hspace*{-0.5mm}
\Tr\Big(\hspace*{-0.5mm}\mathbf{H}_k(\sum\limits_
{\substack{j\neq k}}^K\hspace*{-0.5mm}\mathbf{W}_j+\mathbf{V})\Big)
\hspace*{-0.5mm}+\hspace*{-0.5mm}\sigma_{\mathrm{IR}_k}^2,\forall k,\notag \\
&&\hspace*{-5mm}\mbox{C2: }\notag\max_{\Delta\mathbf{g}_m\in{\Omega }_m} \frac{\Tr(\mathbf{W}_k\mathbf{G}_m)}{ \Gamma_{\mathrm{tol}}}  \le \Tr(\mathbf{G}_m\mathbf{V})+\sigma_{\mathrm{s}}^2, \forall m, k,\\ \notag
&&\hspace*{-5mm}\mbox{C3: }{\sum_{k=1}^K s_{l,k} R_{\mathrm{B}_k} \le C_{l}^{\mathrm{B}_{\max}}, \forall l,}\\
&&\hspace*{-5mm}\mbox{C4: }P_{\mathrm{C}}^\mathrm{CP}\hspace*{-0.5mm}+\hspace*{-0.5mm}\sum_{l=1}^L \Big\{P_{\mathrm{C}_l}\hspace*{-0.5mm}+\hspace*{-0.5mm} \varepsilon\Big(\sum_{k=1}^K \Tr(\mathbf{W}_k\mathbf{R}_l)\hspace*{-0.5mm}+\hspace*{-0.5mm}\Tr(\mathbf{V}\mathbf{R}_l) \Big)\Big\}\notag\\
&&\hspace*{-5mm}\le \mathbf{1}^T\mathbf{e}^{\mathrm{S}}- (\mathbf{e}^{\mathrm{S}} )^T\mathbf{B}\mathbf{e}^{\mathrm{S}},\notag\\
&&\hspace*{-5mm}\mbox{C6: } \Tr(\mathbf{V}\mathbf{R}_l)+\sum_{k=1}^K\Tr\big(\mathbf{R}_l\mathbf{W}_k\big)\le  P^{\mathrm{T}_{\max}}_l,\,\, \forall l,\notag\\
&&\hspace*{-5mm}\mbox{C7: } \notag \min_{\Delta\mathbf{g}_m\in{\Omega }_k}\,\hspace*{-0.5mm}\mu\Big[\hspace*{-0.5mm}\Tr\Big(\big(\sum_{k=1}^K\mathbf{W}_k\hspace*{-0.5mm}+\hspace*{-0.5mm}\mathbf{V}\big)\mathbf{G}_m\Big)\hspace*{-0.5mm}\Big]\ge P^{\min}_{m},\,\forall m, \notag\\
&&\hspace*{-5mm} \notag\mbox{C5},\quad \mbox{C8},\quad \mbox{C9}, \notag\\
&&\hspace*{-5mm}\mbox{C10: } s_{l,k}\in \{0,1\},\forall k,l,\quad\notag\\
&&\hspace*{-5mm} \mbox{C11: } \Tr(\mathbf{W}_k\mathbf{R}_l)\le s_{l,k} P^{\mathrm{T}_{\max}}_l ,\forall k,l,\notag\\
&&\hspace*{-5mm}\mbox{C12:}\,\, \mathbf{W}_k\succeq \mathbf{0},\,\, \forall k, \quad\,\,\,
\mbox{C13:}\,\, \Rank(\mathbf{W}_k)\le 1,\,\, \forall k,
\end{eqnarray}
 Constraints C12, C13, and $\mathbf{W}_k\in\mathbb{H}^{N_{\mathrm{T}} L},\forall k$, are imposed to guarantee that $\mathbf{W}_k=\mathbf{w}_k\mathbf{w}_k^H$ holds after optimization. On the other hand, C10 and C11 are auxiliary constraints. In particular, constraints C10 and C11 restrict the optimization problem such that $s_{l,k}=1$ must hold when the data of IR $k$ is conveyed to RRH $l$ for information transmission, i.e., $\Tr(\mathbf{W}_k\mathbf{R}_l)>0$. In other words, when $\Tr(\mathbf{W}_k\mathbf{R}_l)>0$, the data of IR $k$ consumes $R_{\mathrm{B}_k}$ bit/s/Hz of the capacity  of backhaul link $l$, cf. C3 in (\ref{eqn:equivalent-binary}). On the other hand, it can be verified that the optimization problems in (\ref{eqn:equivalent-binary}) and (\ref{eqn:cross-layer}) are equivalent in the sense that they share the same optimal solution $\{\mathbf{W}_k,\mathbf{V},\mathbf{e}^{\mathrm{S}}, s_{l,k}\}$. As a result, we focus on the design of an algorithm for solving the non-convex  optimization problem  in (\ref{eqn:equivalent-binary}).

\subsection{Iterative Resource Allocation Algorithm}
In the following, we adopt the generalized Bender's decomposition (GBD) to handle the constraints involving binary optimization variables \cite{JR:Vijay_generalized_Benders}--\nocite{book:non_linear_and_mixed_integer,JR:generalized_Bender's}\cite{book:non_linear_integer}, i.e., C3, C10, and C11. In particular, we decompose the problem in  (\ref{eqn:equivalent-binary}) into two problems, a \emph{primal problem} and a \emph{master problem}.  The primal problem is a non-convex optimization problem when optimization variable $s_{l,k}$ is fixed and solving this problem with respect to $\{\mathbf{W}_k,\mathbf{V},\mathbf{e}^{\mathrm{S}}\}$ yields an upper bound for the optimal value of (\ref{eqn:equivalent-binary}).  The master problem is a mixed-integer linear programming (MILP) with  binary optimization variables $s_{l,k}$ for a fixed value of $\{\mathbf{W}_k,\mathbf{V},\mathbf{e}^{\mathrm{S}}\}$. The solution of the master problem  provides a lower bound for the optimal value of (\ref{eqn:equivalent-binary}). We solve the primal and master problems iteratively until the solutions converge. In the following, we first propose algorithms for solving the primal and master problems in the $i$-th iteration, respectively. Then, we describe the iterative procedure between the master problem and the primal problem.
\subsubsection{Solution of the primal problem in the $i$-th iteration}
For given and fixed input parameters  $s_{l,k}(i)$ obtained from the master problem in the $i$-th iteration, we solve the following primal optimization problem:
\begin{eqnarray}\label{eqn:primal_problem}
&&\hspace*{-15mm} \underset{\mathbf{W}_k,\mathbf{V}\in \mathbb{H}^{N_{\mathrm{T}}L},\mathbf{e}^{\mathrm{S}}}{\mino}\,\, \sum_{k=1}^K\Tr(\mathbf{W}_k)+\Tr(\mathbf{V})\notag\\
\mathrm{s.t.} &&\hspace*{-5mm}\mbox{C1, C2, C4 -- C9, C11 -- C13.}
\end{eqnarray}
We note that constraints C3 and C10 in (\ref{eqn:equivalent-binary}) will be handled by the master problem since they involve only the binary optimization variable $s_{l,k}$. Besides, $s_{l,k}$ is treated as a given constant in (\ref{eqn:primal_problem}) and we minimize the objective function with respect to variables $\{\mathbf{W}_k,\mathbf{V},\mathbf{e}^{\mathrm{S}}\}$. The first step in solving the primal problem in (\ref{eqn:primal_problem}) is to handle the infinitely many constraints in C2 and C7 due to the imperfect CSI. To facilitate the  resource allocation algorithm design, we transform constraints C2 and C7 into linear matrix inequalities (LMIs) via the S-Procedure  \cite{book:convex}. Exploiting \cite{book:convex} it can be shown that the original constraint C2 holds if and only if there exist $\delta_{m,k}\ge 0, m\in\{1,\ldots,M\},k\in\{1,\ldots,K\}$,  such that the following  LMI constraints hold:
\begin{eqnarray}\label{eqn:LMI_C2}
&&\mbox{C2: }\mathbf{S}_{\mathrm{C}_{2_{m,k}}}\Big(\mathbf{W}_k,\mathbf{V},\delta_{m,k}\Big)\\
\hspace*{-0.5mm}&=&\hspace*{-0.5mm}
          \begin{bmatrix}
       \delta_{m,k}\mathbf{\Xi}_m+\mathbf{V} & \hspace*{-1mm}\mathbf{V}\mathbf{\hat g}_m          \\
       \mathbf{\hat g}_m^H \mathbf{V}    & \hspace*{-1mm}-\delta_{m,k}\varepsilon_m^2 +\sigma_{\mathrm{s}}^2+  \mathbf{\hat g}_m^H \mathbf{V} \mathbf{\hat g}_m        \\
           \end{bmatrix}\notag\\
           \hspace*{-0.5mm}&-&\hspace*{-0.5mm}\frac{\mathbf{U}_{\mathbf{g}_m}^H\mathbf{W}_k\mathbf{U}_{\mathbf{g}_m}}{\Gamma_{\mathrm{tol}_m}} \succeq \mathbf{0}, \forall k,\notag
\end{eqnarray}
where $\mathbf{U}_{\mathbf{g}_m}=\Big[\mathbf{I}_{N_{\mathrm{T}}L}\quad\mathbf{\hat g}_m\Big]$. Similarly,  constraint C7 can be equivalently written as
\begin{eqnarray}\label{eqn:LMI_C7}&&
\mbox{C7: }\mathbf{S}_{\mathrm{C}_{7_m}}\Big(\mathbf{W}_k,\mathbf{V}, \mathbf{\nu}_m\Big)\\
&=&\notag
           \begin{bmatrix}
       \nu_m\mathbf{\Xi}_m+\mathbf{V}& \mathbf{V}\mathbf{\hat g}_m          \\
       \mathbf{\hat g}_m^H \mathbf{V}
        & -\nu_m\varepsilon_m^2 -\frac{P_{\min_m}}{\mu}  +  \mathbf{\hat g}_m^H \mathbf{V} \mathbf{\hat g}_m        \\
           \end{bmatrix}\notag\\
          & +& \sum_{k=1}^K \mathbf{U}_{\mathbf{g}_m}^H\mathbf{W}_k\mathbf{U}_{\mathbf{g}_m}\succeq \mathbf{0}, \forall m,\notag
\end{eqnarray}
for $\nu_m\ge 0, m\in\{1,\ldots,M\}$. Now, constraints C2 and C7 involve only a finite number of constraints which facilitates the resource allocation algorithm design. As a result, we can rewrite the primal problem as:
 \begin{eqnarray}\label{eqn:equivalent}\notag
&& \hspace*{-5mm}\underset{\underset{\mathbf{e}^{\mathrm{S}},\boldsymbol \delta, \boldsymbol \nu }{\mathbf{W}_k,\mathbf{V}\in\mathbb{H}^{N_{\mathrm{T}}L}}}{\mino} \,\, \sum_{k=1}^K\Tr(\mathbf{W}_k)+\Tr(\mathbf{V})\\
 &&\hspace*{-15mm}\mathrm{s.t. }\,\mbox{C1, C4, C5, C6, C8, C9, C12,}\notag \\
&&\hspace*{-10mm}\mbox{C2: }\notag \mathbf{S}_{\mathrm{C}_{2_{m,k}}}\big(\mathbf{W}_k,\mathbf{V},\delta_{m,k}\big)\succeq\zero, \forall m, k,\notag\\
&&\hspace*{-10mm}\mbox{C7: }\mathbf{S}_{\mathrm{C}_{7_m}}\Big(\mathbf{W}_k,\mathbf{V}, \mathbf{\nu}_m\Big)\succeq\zero,\,\, \forall m, \notag\\
&&\hspace*{-10mm} \notag \mbox{C11: } \Tr(\mathbf{W}_k\mathbf{R}_l)\le s_{l,k}(i) P^{\mathrm{T}_{\max}}_l ,\forall k,l, \,\,\,\\
&&\hspace*{-10mm}\mbox{C13:}\,\, \Rank(\mathbf{W}_k)\le 1,\,\, \forall k,\notag\\
 &&\hspace*{-10mm}\mbox{C14:}\,\, \delta_{m,k},\nu_m\ge 0,\,\, \forall m,k,
\end{eqnarray}
where $\boldsymbol \delta$ and $\boldsymbol \nu$ are auxiliary  optimization variable vectors, whose elements $\delta_{m,k}\ge0, m\in\{1,\ldots,M\}$, $k\in\{1,\ldots,K\}$, and $\nu_m\ge0, m\in\{1,\ldots,M\}$, were introduced in (\ref{eqn:LMI_C2}) and (\ref{eqn:LMI_C7}), respectively.
Then, we relax constraint $\mbox{C13: }\Rank(\mathbf{W}_k)\le1$ by removing it from the problem formulation, such that the considered problem becomes a convex semidefinite program (SDP).  We note that the relaxed problem of (\ref{eqn:equivalent}) can be solved efficiently by convex programming numerical solvers such as  CVX \cite{website:CVX}. If the matrices $\mathbf{W}_k$ obtained from the relaxed problem  (\ref{eqn:equivalent}) are rank-one matrices for all IRs, $k\in\{1,\ldots,K\}$, then  the problem in (\ref{eqn:equivalent}) and its relaxed version share the same optimal solution and the same optimal objective value. Otherwise, the optimal objective value of the relaxed version of (\ref{eqn:equivalent})  serves as a lower bound for the objective value of (\ref{eqn:equivalent}) since a larger feasible solution set is considered.

Now, we study the tightness of the adopted SDP relaxation. As the  SDP relaxed optimization problem in (\ref{eqn:equivalent}) satisfies Slater's constraint qualification and is jointly convex with respect to the optimization variables, strong duality holds and thus solving the dual problem is equivalent to solving (\ref{eqn:equivalent}). For formulating the dual problem, we first define the Lagrangian
 of the relaxed version of (\ref{eqn:equivalent}) which can be expressed as
\begin{eqnarray}
&&{\cal L}\Big(\mathbf{W}_k,\mathbf{V},\mathbf{e}^{\mathrm{S}},\boldsymbol \delta, \boldsymbol \nu,  s_{l,k}(i),\boldsymbol \Phi\Big)\\
=&&f_0(\mathbf{W}_k,\mathbf{V})\hspace*{-0.5mm}+\hspace*{-0.5mm}f_1(\mathbf{W}_k,\mathbf{V},\mathbf{e}^{\mathrm{S}},\boldsymbol \delta, \boldsymbol \nu,  \boldsymbol \Phi)\notag\\
 \hspace*{-0.5mm}+\hspace*{-0.5mm}&& f_2(\mathbf{W}_k, s_{l,k}(i),\boldsymbol \Phi),\,\quad \mbox{where}\notag\\
&&\hspace*{-10mm}\label{eqn:f0}f_0(\mathbf{W}_k,\mathbf{V})=\sum_{k=1}^K\Tr(\mathbf{W}_k)+\Tr(\mathbf{V}),\\
&&\hspace*{-10mm}\label{eqn:f1}f_1(\mathbf{W}_k,\mathbf{V},\mathbf{e}^{\mathrm{S}},\boldsymbol \delta, \boldsymbol \nu,\boldsymbol \Phi)=-\Tr(\mathbf{Y}\mathbf{V})-\sum_{k=1}^K\Tr(\mathbf{Z}_k\mathbf{W}_k)\notag\\
&&\hspace*{-10mm}+
\sum_{n=1}^{L+1}\tau_n(\boldsymbol{\theta}_n^T\mathbf{e}^{\mathrm{S}}- E^{\max}_n)\notag\\
&&\hspace*{-10mm}-  \sum_{m=1}^{M}\sum_{k=1}^{K}\Tr\Big(\mathbf{S}_{\mathrm{C}_{2_{m,k}}}
\big(\mathbf{W}_k,\mathbf{V},\delta_{m,k}\big)\mathbf{D}_{\mathrm{C}_{2_{m,k}}}\Big)\notag\\
&&\hspace*{-10mm}-\sum_{m=1}^{M}
\Tr\Big(\mathbf{S}_{\mathrm{C}_{7_m}}\Big(\mathbf{W}_k,\mathbf{V}, \mathbf{\nu}_m\Big)\mathbf{D}_{\mathrm{C}_{7_m}}\Big)-\sum_{n=1}^{L+1}(\boldsymbol{\theta}_n^T\mathbf{e}^{\mathrm{S}})\chi_n\ \notag\\
&&\hspace*{-10mm}+
\sum_{k=1}^K\alpha_k\Big[-\frac{\Tr(\mathbf{H}_k\mathbf{W}_k)}{\Gamma_{\mathrm{req},k}}+
\Tr\Big(\mathbf{H}_k(\sum\limits_
{\substack{j\neq k}}^K\mathbf{W}_j+\mathbf{V})\Big)+\sigma_{\mathrm{IR}_k}^2\Big]
\notag\end{eqnarray}
\begin{eqnarray}
&&\hspace*{-5mm}+\varrho\Big(P_{\mathrm{C}}^\mathrm{CP}+\sum_{l=1}^L \Big\{P_{\mathrm{C}_l}+ \varepsilon\Big(\sum_{k=1}^K\Tr(\mathbf{W}_k\mathbf{R}_l)+\Tr(\mathbf{V}\mathbf{R}_l) \Big)\Big\}\notag\\
&&\hspace*{-5mm}- \mathbf{1}^T\mathbf{e}^{\mathrm{S}}+ (\mathbf{e}^{\mathrm{S}} )^T\mathbf{B}\mathbf{e}^{\mathrm{S}}\Big)-
\sum_{k=1}^K\sum_{m=1}^M\delta_{m,k}\lambda_{m,k}-\sum_{m=1}^M\nu_{m}\theta_{m}\notag\\
&&\hspace*{-5mm}+\sum_{l=1}^L\gamma_{l} \Big(\Tr(\mathbf{V}\mathbf{R}_l)+\sum_{k=1}^K\Tr\big(\mathbf{R}_l\mathbf{W}_k\big)-  P^{\mathrm{T}_{\max}}_l\Big), \,\, \mbox{and}\\
&&\hspace*{-5mm}\label{eqn:f2}f_2(\mathbf{W}_k, s_{l,k}(i),\boldsymbol \Phi)\notag\\
&&\hspace*{-9mm}=\sum_{k=1}^K\sum_{l=1}^L\beta_{k,l}\Big(\Tr(\mathbf{W}_k\mathbf{R}_l)- s_{l,k}(i) P^{\mathrm{T}_{\max}}_l\Big),
\end{eqnarray}
respectively. Here, $\boldsymbol \Phi=\{\mathbf{D}_{\mathrm{C}_{2_{m,k}}},\mathbf{D}_{\mathrm{C}_{7_m}},\mathbf{Y},\mathbf{Z}_k,\alpha_k,\varrho,\tau_n,$ $\chi_n,\gamma_{l},\beta_{k,l},\lambda_{m,k},\theta_m\}$ is a collection of dual variables; $\mathbf{D}_{\mathrm{C}_{2_{m,k}}}$, $\mathbf{D}_{\mathrm{C}_{7_m}}$, $\mathbf{Y}$, and $\mathbf{Z}_k$ are the dual variable matrices for constraints  C2, C7, C9, and C12, respectively; $\alpha_k$, $\varrho$, $\tau_n$, $\chi_n$,  $\gamma_l$, $\beta_{k,l}$, and  $\lambda_{m,k},\theta_m$ are the scalar dual variables for constraints C1, C4, C5, C6, C8, C11, and  C14, respectively.
Function $f_0(\mathbf{W}_k,\mathbf{V})$  in (\ref{eqn:f0}) is the objective function of the SDP relaxed version of (\ref{eqn:equivalent}); $f_1(\mathbf{W}_k,\mathbf{V},\mathbf{e}^{\mathrm{S}},\boldsymbol \delta, \boldsymbol \nu,  \boldsymbol \Phi)$ in (\ref{eqn:f1}) is a function involving only continuous optimization variables and dual variables;
$f_2(\mathbf{W}_k, s_{l,k}(i),\boldsymbol \Phi)$ in (\ref{eqn:f2}) is a function involving continuous optimization variables, dual variables, and  binary optimization variable $s_{l,k}(i)$. These functions are defined here for notational simplicity and will be exploited for facilitating the presentation of the solutions for both the primal problem and the master problem.

The dual problem of the relaxed SDP optimization problem in (\ref{eqn:equivalent}) is given by
\begin{equation}\hspace*{-0mm}\label{eqn:dual}
\underset{\boldsymbol \Phi\succeq \zero}{\maxo} \,\underset{\underset{\mathbf{e}^{\mathrm{S}},\boldsymbol \delta, \boldsymbol \nu }{\mathbf{W}_k,\mathbf{V}\in\mathbb{H}^{N_{\mathrm{T}}L}}}{\mino} \,{\cal L}\Big(\mathbf{W}_k,\mathbf{V},\mathbf{e}^{\mathrm{S}},\boldsymbol \delta, \boldsymbol \nu,  s_{l,k}(i),\boldsymbol \Phi\Big).
\end{equation}
 We define $\boldsymbol \Theta(i)=\{\mathbf{W}_k^*,\mathbf{V}^*,\mathbf{e}^{\mathrm{S}*},\boldsymbol \delta^*, \boldsymbol \nu^* \}$ and $\boldsymbol \Phi(i)=\{\boldsymbol \Phi^* \}$ as the optimal primal solution and the optimal dual solution of the SDP relaxed problem in (\ref{eqn:equivalent}) in the $i$-th iteration.

{In the following, we introduce a  theorem inspired by \cite{JR:rui_zhang} revealing the tightness of the SDP relaxation adopted in  (\ref{eqn:equivalent}). Let $\mathbf{C}_k=\mathbf{I}_{N_\mathrm{T}L}+\sum_{m=1}^M\mathbf{U}_{\mathbf{g}_m}(\frac{\mathbf{D}_{\mathrm{C}_{2_{m,k}}} }{{\Gamma}_{\mathrm{req}_k}} -\mathbf{D}_{\mathrm{C}_{7_{m}}})\mathbf{U}_{\mathbf{g}_m}^H+\sum_{j\ne k} \mathbf{H}_j\alpha_j+\sum_{l=1}^L\mathbf{R}_l ( \varrho\varepsilon+\gamma_l+\beta_{l,k})$ and $\Rank(\mathbf{C}_k)=r_k$. In addition, we denote the orthonormal basis of the null space of $\mathbf{C}_k$ as $\mathbf{\mathbf{\Upsilon}}_k\in\mathbb{C}^{N_{\mathrm{T}}L\times (N_{\mathrm{T}}L-r_{k})}$, and ${\boldsymbol \phi}_{\omega_{k}}\in \mathbb{C}^{N_{\mathrm{T}}L\times 1}$,  $1\le \omega_{k}\le N_{\mathrm{T}}L-r_{k}$, denotes the $\omega_{k}$-th column  of $\mathbf{\Upsilon}_{k}$. Hence, $\mathbf{C}_{k}\mathbf{\Upsilon}_{k}=\zero$ and $\Rank(\mathbf{\Upsilon}_{k})=N_{\mathrm{T}}L-r_{k}$.

\begin{Thm}\label{thm:rankone_condition}For ${\Gamma}_{\mathrm{req}_k}>0$ and $\Gamma_{\mathrm{tol}}>0$, the optimal primal and dual solutions of the SDP relaxed version of (\ref{eqn:equivalent}), denoted by $\boldsymbol \Theta^*=\{\mathbf{W}_k^*,\mathbf{V}^*,\mathbf{e}^{\mathrm{S}*},\boldsymbol \delta^*, \boldsymbol \nu^*\}$  and $\boldsymbol \Phi^*=\{\mathbf{D}_{\mathrm{C}_{2_{m,k}}}^*,\mathbf{D}_{\mathrm{C}_{7_m}}^*,$ $\mathbf{Y}^*,\mathbf{Z}_k^*,\alpha_k^*,\varrho^*,\tau_n^*,\chi_n^*,\gamma_{l}^*,\beta_{k,l}^*,$ $\lambda_{m,k}^*,\theta_m^*\}$, respectively, satisfy the following conditions:
 \begin{enumerate}
 \item The optimal beamforming matrix $\mathbf{W}_k^*$ can be expressed as
 \begin{eqnarray}
     \mathbf{W}^*_{k}=\sum_{\omega_{k}=1}^{N_{\mathrm{T}}L-r_{k}}\psi_{\omega_{k}}{\boldsymbol \phi}_{\omega_{k}}  {\boldsymbol \phi}_{\omega_{k}} ^H  + \underbrace{f_{k}\mathbf{u}_{k}\mathbf{u}^H_{k}}_{\mbox{rank-one}},
     \end{eqnarray}
where variables $\psi_{\omega_{k}}\ge0, \forall \omega_{k}\in\{1,\ldots,N_{\mathrm{T}}L-r_{k}\},$ and $f_{k}>0$ are positive scalars and $\mathbf{u}_{k}\in \mathbb{C}^{N_{\mathrm{T}}L\times 1}$, $\norm{\mathbf{u}_{k}}=1$, such that $\mathbf{u}^H_{k}\mathbf{\Upsilon}_{k}=\zero$.
 \item At the optimal solution, the null space of matrix $\mathbf{C}_k$, denoted as $\mathbf{\Upsilon}_{k}^*$, satisfies the following equality:
     \begin{eqnarray}\mathbf{H}_k\mathbf{\Upsilon}_{k}^*=\zero.\end{eqnarray}
      \item If  $\exists k:\Rank(\mathbf{W}^*_k)>1$, i.e., $\psi_{\omega_{k}}>0$, then we can construct another solution of (18), denoted by  $\{\mathbf{\overline W}_k ,\mathbf{\overline V},\mathbf{\overline e}^{\mathrm{S}}, \boldsymbol{\overline\delta},  \boldsymbol{\overline\nu}\}$, which not only achieves the same objective value as $\boldsymbol \Theta^*$, but also admits a rank-one beamforming matrix, i.e.,  $\Rank(\mathbf{\overline W}_k)=1,\forall k$. The new optimal solution for the primal problem in the $i$-th iteration is given by
\begin{eqnarray}\label{eqn:rank-one-structure}\notag\mathbf{\overline W}_{k}\hspace*{-2mm}&=&\hspace*{-2mm}f_{k}\mathbf{u}_{k}\mathbf{u}^H_{k}=\mathbf{W}^*_{k}-\sum_{\omega_{k}=1}^{N_{\mathrm{T}}L-r_{k}} \psi_{\omega_{k}} {\boldsymbol \phi}_{\omega_{k}} {\boldsymbol \phi}_{\omega_{k}}^H,\\
 \mathbf{\overline V}\hspace*{-2mm}&=&\hspace*{-2mm}\mathbf{ V}^*+\sum_{\omega_{k}=1}^{N_{\mathrm{T}}L-r_{k}} \psi_{\omega_{k}} {\boldsymbol \phi}_{\omega_{k}} {\boldsymbol \phi}_{\omega_{k}}^H,\quad \mathbf{\overline e}^{\mathrm{S}}= \mathbf{e}^{\mathrm{S}*},\notag\\
  \boldsymbol{\overline\delta}\hspace*{-2mm}&=&\hspace*{-2mm}\boldsymbol \delta^*,\quad   \boldsymbol{\overline\nu}=\boldsymbol \nu^*,\quad
\end{eqnarray}
 \end{enumerate}
with $\Rank(\mathbf{\overline W}_{k})=1,\forall k\in\{1,\ldots,K\}$, where $f_{k}$ and $ \psi_{\omega_{k}}$ can be easily
found  by  applying above $3$ conditions to the relaxed version of (\ref{eqn:equivalent}) and solving the resulting convex
optimization problem for $f_{k}$ and $ \psi_{\omega_{k}}$.\end{Thm}}

\emph{\quad Proof: } The  proof of Theorem 1 closely follows the proof of  \cite[Proposition 4.1]{JR:rui_zhang} and is omitted here due to page limitation.\qed

In other words, by applying Theorem \ref{thm:rankone_condition}, the optimal solution of  the primal problem is obtained in each iteration. Besides, from the numerical solver, the dual variables  corresponding to the constraints in (\ref{eqn:equivalent}), i.e., $\boldsymbol \Phi$, are obtained together with the primal solution $\boldsymbol \Theta$. This information is used as an input to the master problem.

{
If problem (\ref{eqn:equivalent}) is infeasible for a given binary variable $s_{l,k}(i)$, then we formulate an $l_1$-minimization problem and use the corresponding dual variables and the optimal primal variables as the input to the master problem for the next iteration \cite{book:non_linear_and_mixed_integer}. The $l_1$-minimization problem is given as:
 \begin{eqnarray}\label{eqn:FP}\notag
&& \hspace*{0mm}\underset{\underset{\mathbf{e}^{\mathrm{S}},\boldsymbol \delta, \boldsymbol \nu }{\mathbf{W}_k,\mathbf{V}\in\mathbb{H}^{N_{\mathrm{T}}L}}}{\mino}\,\, \sum_{k=1}^K\sum_{l=1}^L \alpha_{l,k}\\
\hspace*{0mm} &&\hspace*{-15mm}\mathrm{s.t.}\,\,\mbox{C1, C2, C4 -- C9, C11, C12, C14,}\notag \\
&&\hspace*{-8mm}\mbox{C11: } \Tr(\mathbf{W}_k\mathbf{R}_l)\le s_{l,k}(i) P^{\mathrm{T}_{\max}}_l + \alpha_{l,k},\forall k,l,\notag\\
&&\hspace*{-8mm}\mbox{C15:}\,\, \alpha_{l,k}\ge 0, \forall l,k.
\end{eqnarray}
The $l_1$-minimization problem is a convex optimization problem and can be solved by standard convex programming solvers.
The optimal value of the $l_1$-minimization problem measures the aggregated violations  of the constraints for a given $s_{l,k}(i)$. We adopt a similar notation as in (\ref{eqn:equivalent})  to denote the dual variables with respect to  constraints C1, C2, C4 -- C9, C11, C12, and C14 in (\ref{eqn:FP}). In particular, these variables are defined as:
$\boldsymbol{\widetilde\Phi}(i)=\{\mathbf{\widetilde D}_{\mathrm{C}_{2_{m,k}}},\mathbf{\widetilde D}_{\mathrm{C}_{7_m}},\mathbf{\widetilde Z}_k,\mathbf{\widetilde Y},$ $\widetilde\alpha_k,\widetilde\varrho,\widetilde\tau_n,\widetilde\chi_n,\widetilde\gamma_{l},
\widetilde\beta_{k,l},\widetilde\lambda_{m,k},\widetilde\theta_m\}$. Also, the
solution for the $l_1$-minimization problem in (\ref{eqn:FP}) is denoted as  $\mathbf{\widetilde \Theta}(i)=\{\mathbf{W}_k,\mathbf{V},
\mathbf{e}^{\mathrm{S}},\boldsymbol \delta, \boldsymbol \nu\}$. The primal and dual solutions of the $l_1$-minimization problem are used to generate a \emph{feasibility cut} which separates the current infeasible solution from the search space in the master problem.}
\subsubsection{Solution of the master problem in the $i$-th iteration}
For notational simplicity, we define $\cal F$ and $\cal I$ as the sets of all iteration indices at which the primal problem is feasible and infeasible, respectively. Then, we formulate the master problem which utilizes the solutions of (\ref{eqn:equivalent}) and (\ref{eqn:FP}). The master problem in the $i$-th iteration is given as follows:
\begin{subequations}\label{eqn:master_problem}
\begin{eqnarray}\label{eqn:master_problem_objective}
&&\hspace*{5mm}\underset{\mu,\, s_{l,k}}{\mino}\,\, \mu\\
\hspace*{5mm}\mathrm{s.t.} &&\mu \ge  \xi(\boldsymbol \Phi(t),s_{l,k}),   t\in\{1,\ldots,i\} \cap   \cal F,\label{eqn:supporting_plane_constraint1}\\
&&0\ge  \overline{ \xi}( \boldsymbol{\widetilde\Phi}(t),s_{l,k}),  t\in\{1,\ldots,i\} \cap  \cal I,\label{eqn:supporting_plane_constraint2}\\
&&\hspace*{-22mm}\mbox{C3: }\sum_{k=1}^K s_{l,k} R_{\mathrm{B}_k} \le C_{l}^{\mathrm{B}_{\max}}, \forall l, \quad\mbox{C10: } s_{l,k}\in \{0,1\},
\end{eqnarray}
\end{subequations}
where $s_{l,k}$ and $\mu$ are optimization variables for the master problem and
\begin{eqnarray}\label{eqn:opt_master_1}\notag
\xi( \boldsymbol{\Phi}(t),s_{l,k})\hspace*{-1.5mm}&=&\hspace*{-1.5mm} \underset{\underset{\mathbf{e}^{\mathrm{S}},\boldsymbol \delta, \boldsymbol \nu }{\mathbf{W}_k,\mathbf{V}\in\mathbb{H}^{N_{\mathrm{T}}L}}}{\mino}\,\, \Big\{f_0(\mathbf{W}_k,\mathbf{V})\\
&&\hspace*{-25mm}+f_1(\mathbf{W}_k,\mathbf{V},\mathbf{e}^{\mathrm{S}},\boldsymbol \delta, \boldsymbol \nu,  \boldsymbol \Phi(t)) + f_2(\mathbf{W}_k, s_{l,k},\boldsymbol \Phi(t))\Big\},\\ \notag
\overline\xi( \boldsymbol{\widetilde\Phi}(t),s_{l,k})\hspace*{-1.5mm}&=& \hspace*{-1.5mm}\underset{\underset{\mathbf{e}^{\mathrm{S}},\boldsymbol \delta, \boldsymbol \nu }{\mathbf{W}_k,\mathbf{V}\in\mathbb{H}^{N_{\mathrm{T}}L}}}{\mino}\,\, \Big\{f_1(\mathbf{W}_k,\mathbf{V},\mathbf{e}^{\mathrm{S}},\boldsymbol \delta, \boldsymbol \nu,  \boldsymbol{\widetilde \Phi}(t)) \\
&&\hspace*{-25mm}+ f_2(\mathbf{W}_k, s_{l,k},\boldsymbol{\widetilde \Phi}(t))\Big\}.\label{eqn:opt_master_2}
\end{eqnarray}
Equations (\ref{eqn:opt_master_1}) and (\ref{eqn:opt_master_2}) represent two different inner minimization problems inside the master problem.  In particular, $\mu\ge \xi( \boldsymbol{\Phi}(t),s_{l,k}), t\in\{1,\ldots,i\} \cap  \cal F$ and $0\ge\overline\xi( \boldsymbol{\widetilde\Phi}(t),s_{l,k}), t\in\{1,\ldots,i\} \cap  \cal I,$ denote the sets of hyperplanes spanned by the \emph{optimality cut} and \emph{feasibility cut} from the first to the  $i$-th iteration, respectively. The two different types of cuts are exploited to reduce the search region for the global optimal solution. Besides, both $\xi( \boldsymbol{\Phi}(t),s_{l,k})$ and $\overline\xi( \boldsymbol{\widetilde\Phi}(t),s_{l,k})$ are also functions  of $s_{l,k}$ which is the optimization variable of the outer minimization in (\ref{eqn:master_problem}).

Now, we introduce the following proposition for the solutions of the inner minimization problems.

\begin{proposition}\label{thm:supporting_plane}
The solutions of (\ref{eqn:opt_master_1}) and (\ref{eqn:opt_master_2}) for index $t\in\{1,\ldots,i\}$  are  the solutions of (\ref{eqn:equivalent}) and (\ref{eqn:FP}) in the $t$-th iteration, respectively.
\end{proposition}

\emph{\quad Proof: } Please refer to the Appendix for a proof of Proposition \ref{thm:supporting_plane}.

By substituting $\mathbf{\Theta}(t)$ and $\mathbf{\widetilde\Theta}(t)$ into (\ref{eqn:opt_master_1}) and (\ref{eqn:opt_master_2}), respectively, the master problem is a standard MILP which can be solved by using standard numerical solvers for MILPs such as Mosek \cite{JR:Mosek} and Gurobi \cite{JR:Gurobi}. We note that the objective value of (\ref{eqn:master_problem}), i.e., (\ref{eqn:master_problem_objective}), is a monotonically non-decreasing function with respect to the number of iterations as an additional constraint is imposed to the master problem in each additional
 iteration.

\subsubsection{Overall algorithm} The overall iterative resource allocation algorithm is summarized in Table \ref{table:algorithm}.
\begin{table}[t]\vspace*{-0cm}\caption{Optimal Iterative Resource Allocation Algorithm based on GBD.\vspace*{-0.2cm}}\label{table:algorithm}

\renewcommand\thealgorithm{}
\begin{algorithm} [H]                
\caption{Generalized Bender's Decomposition}          
\label{alg1}                           
\begin{algorithmic} [1]
      \STATE Initialize the maximum number of iterations $L_{\max}$ and a small constant $\kappa\rightarrow 0$
\STATE Set iteration index $i=0$ and start with  random  values $s_{l,k}(i),\forall k,l$

\REPEAT [Loop]

\STATE Solve  (\ref{eqn:equivalent}) according to Theorem 1 for a given set of $s_{l,k}(i)$

\IF{(\ref{eqn:equivalent}) is feasible}

\STATE Obtain an intermediate resource allocation policy $\mathbf{\Theta}(i)=\{\mathbf{W}_k',\mathbf{V}',\mathbf{e}^{\mathrm{S}'},\boldsymbol \delta', \boldsymbol \nu'\}$, the corresponding Lagrange multiplier set $\boldsymbol\Phi(i)$, and an intermediate objective value $f_0'$

\STATE Update the upper bound $\mathrm{UB}(i)={\min} \{\mathrm{UB}(i-1), f_0'\} $. If $\mathrm{UB}(i)=f_0'$, set the current optimal policy $\boldsymbol \Theta_{\mathrm{current}}=\boldsymbol\Theta(i)$, $s_{\mathrm{current}}= s_{l,k}(i)$

\ELSE
\STATE  Solve the feasibility problem in (\ref{eqn:FP}) and obtain an intermediate resource allocation policy $\mathbf{\widetilde \Theta}(i)=\{\mathbf{W}_k',\mathbf{V}',\mathbf{e}^{\mathrm{S}'},\boldsymbol \delta', \boldsymbol \nu'\}$ and the corresponding Lagrange multiplier set  $\boldsymbol{\widetilde\Phi}(i)$
 \ENDIF

\STATE Solve the master problem in (\ref{eqn:master_problem}) for $s_{l,k}$, save $s_{l,k}(i+1)=s_{l,k}$,  and obtain the $i$-th lower bound, i.e., $\mathrm{LB}(i)$

\IF{$\abs{\mathrm{LB}(i)-\mathrm{UB}(i)}\le \kappa$}
\STATE
Global optimal = \TRUE, \textbf{return} $\{\mathbf{W}_k^*,\mathbf{V}^*,\mathbf{e}^{\mathrm{S}*},\boldsymbol \delta^*, \boldsymbol \nu^*, s_{l,k}^*\}=\{\boldsymbol \Theta_{\mathrm{current}}, \,s_{\mathrm{current}}\}$
\ELSE
\STATE $i=i+1$
 \ENDIF
\UNTIL{ $i=L_{\max}$}

\end{algorithmic}
\end{algorithm}
\end{table}
The  algorithm is implemented by a repeated loop.  We first set the iteration index $i$ to zero and initialize the binary variables $s_{l,k}(i)$. In the $i$-th iteration, we solve the problem in (\ref{eqn:equivalent}) by  Theorem 1.  If the problem is feasible (lines 6 -- 7),  then we obtain an intermediate resource allocation policy $\mathbf{\Theta}(i)$, the corresponding Lagrange multiplier set $\boldsymbol\Phi(i)$, and an intermediate objective value $f_0'$. Both  $\mathbf{\Theta}(i)$ and $\boldsymbol\Phi(i)$ are used to generate an \emph{optimality cut} in the master problem. Besides, we update the performance upper bound $\mathrm{UB}(i)$ and the current optimal resource allocation policy when the current objective value is the lowest compared to those in all previous iterations.    If the problem is infeasible (lines 9 -- 10), then we solve the $l_1$-minimization problem in (\ref{eqn:FP}) and obtain an intermediate resource allocation policy $\mathbf{\widetilde \Theta}(i)$ and the corresponding Lagrange multiplier set $\boldsymbol{\widetilde \Phi}(i)$. This information will be used to generate an \emph{infeasibility  cut} in the master problem. Then, we solve the master problem  based on $\mathbf{\widetilde \Theta}(t)$ and $\mathbf{\Theta}(i)$, $t\in\{1,\ldots,i\}$,   using a standard MILP numerical solver. The objective value of the master problem in each iteration serves as a system performance lower bound to the original optimization problem in (\ref{eqn:equivalent}) \cite{book:non_linear_and_mixed_integer,JR:generalized_benders}. In the $i$-th iteration, when the difference between the $i$-th lower bound and the $i$-th upper bound is less than a predefined threshold $\kappa$ (lines 12 -- 14), the algorithm stops.  We note that the convergence of the proposed iterative algorithm  to the global optimal solution of (\ref{eqn:equivalent}) in a finite number of iterations is ensured even if $\kappa=0$, provided that the master and primal problems can be solved in each iteration \cite[Theorem 6.3.4]{book:non_linear_and_mixed_integer}. We note that the optimal resource allocation
algorithm has a non-polynomial time computational complexity. Please refer to the simulation section for the illustration of the convergence of the proposed optimal algorithm.

\subsection{Suboptimal Resource Allocation Algorithm Design}
The iterative resource allocation algorithm proposed in the last section leads to the optimal system performance.  However, the algorithm has a non-polynomial time computational complexity since it needs to solve an MILP master problem in each iteration. In this section, we propose a suboptimal resource allocation algorithm which has a polynomial time computational complexity. We start the suboptimal resource allocation algorithm design by focusing on the reformulated optimization problem in (\ref{eqn:equivalent-binary}).
\subsubsection{Problem reformulation via difference of convex functions programming}
The major obstacle in solving (\ref{eqn:equivalent-binary}) is to handle the binary constraint. In fact, constraint C10 is equivalent to
\begin{eqnarray}
\mbox{C10a: }&& 0 \le s_{l,k}\le 1 \quad \mbox{and}\notag\\
 \mbox{C10b: }&& \sum_{l=1}^L \sum_{k=1}^K  s_{l,k} -\sum_{l=1}^L \sum_{k=1}^K  s_{l,k}^2 \le 0,
\end{eqnarray}
where optimization variable $s_{l,k}$ in C10a is a continuous value between zero and one and   C10b is the difference of two convex functions. By using the SDP relaxation approach as in the optimal resource allocation algorithm, we can rewrite the optimization problem as
 \begin{eqnarray}\label{eqn:equivalent-approx}\notag
&&\hspace*{-10mm}\underset{\underset{s_{l,k},\mathbf{e}^{\mathrm{S}},\boldsymbol \delta, \boldsymbol \nu }{\mathbf{W}_k,\mathbf{V}\in\mathbb{H}^{N_{\mathrm{T}}L}}}{\mino}\,\, \sum_{k=1}^K\Tr(\mathbf{W}_k)+\Tr(\mathbf{V})\\
\hspace*{-35mm}\mathrm{s.t.} &&\hspace*{0mm}\mbox{C1 -- C9, C10a, C10b, C11, C12, C14}.
\end{eqnarray}
On the other hand, for a large constant value of $\phi\gg1$, we can follow a similar approach as in \cite{JR:DC_programming} to show that the optimization problem in (\ref{eqn:equivalent-approx}) is equivalent to  the following problem:
 \begin{eqnarray}\label{eqn:equivalent-approx2}\notag
&& \hspace*{-12mm}\underset{\underset{s_{l,k},\mathbf{e}^{\mathrm{S}},\boldsymbol \delta, \boldsymbol \nu }{\mathbf{W}_k,\mathbf{V}\in\mathbb{H}^{N_{\mathrm{T}}L}}}{\mino}\, \sum_{k=1}^K\Tr(\mathbf{W}_k)\hspace*{-0.5mm}+\hspace*{-0.5mm}\Tr(\mathbf{V})\hspace*{-0.5mm}+\hspace*{-0.5mm}\phi\Big( \sum_{l=1}^L \sum_{k=1}^K  (s_{l,k} - s_{l,k}^2)\Big)\\
\mathrm{s.t.} &&\hspace*{10mm}\mbox{C1 -- C9, C10a, C11, C12, C14},
\end{eqnarray}
where $\phi$ acts as a large penalty factor for penalizing the objective function for any $s_{l,k}$ that is not equal to $0$ or $1$. We note that the constraints in (\ref{eqn:equivalent-approx2}) span a convex set which allows the development of an efficient resource allocation algorithm. The problem in  (\ref{eqn:equivalent-approx2})  is known as difference of convex  functions (d.c.) programming due to the convexity of $g(s_{l,k})=\sum_{l=1}^L \sum_{k=1}^K  s_{l,k}^2$. Here, we can apply the successive convex approximation\footnote{ This method is also known as majorization minimization. There are infinitely many of d.c. representations for (\ref{eqn:equivalent-approx}) leading to different successive convex programs. Please refer to \cite{JR:DC_programming} for a more detailed discussion for d.c. programming.} to obtain a locally optimal solution of  (\ref{eqn:equivalent-approx2})  \cite{book:SCA_convergence}.

\begin{table}[t]\vspace*{-0cm}\caption{Suboptimal Iterative Resource Allocation Algorithm}\label{table:algorithm2}

\renewcommand\thealgorithm{}
\begin{algorithm} [H]                    
\caption{Successive Convex Approximation}          
\label{alg1}                           
\begin{algorithmic} [1]
\STATE Initialize the maximum number of iterations $L_{\max}$, penalty factor $\phi\gg 1$, iteration index $i=0$, and $s_{l,k}^{(i)}$
\REPEAT [Loop]
\STATE Solve  (\ref{eqn:equivalent-approx3}) for a given $s_{l,k}^{(i)}$ and obtain the intermediate resource allocation policy $\{\mathbf{W}_k',\mathbf{V}',\mathbf{e}^{\mathrm{S}'},s_{l,k}'\}$
\STATE Set $s_{l,k}^{(i+1)}=s_{l,k}', i=i+1$
\UNTIL{ Convergence or $i=L_{\max}$}

\end{algorithmic}
\end{algorithm}

\end{table}

\subsubsection{Iterative suboptimal algorithm}
The first step is to linearize  the convex function $g(s_{l,k})$. Since  $g(s_{l,k})$ is a differentiable convex function, then the following inequality \cite{book:convex}
\begin{eqnarray}
g(s_{l,k})\ge g(s_{l,k}^{(i)}) +\nabla_{s_{l,k}}g(s_{l,k}^{(i)}) (s_{l,k}-s_{l,k}^{(i)})
\end{eqnarray}
always holds for any feasible point $s_{l,k}^{(i)}$. As a result, for a given value  $s_{l,k}^{(i)}$, the optimal value of the optimization problem,
 \begin{eqnarray}\label{eqn:equivalent-approx3}\notag
&&\hspace*{-0mm} \underset{\underset{s_{l,k},\mathbf{e}^{\mathrm{S}},\boldsymbol \delta, \boldsymbol \nu }{\mathbf{W}_k,\mathbf{V}\in\mathbb{H}^{N_{\mathrm{T}}L}}}{\mino}\,\, \sum_{k=1}^K\Tr(\mathbf{W}_k)\hspace*{-0.5mm}+\hspace*{-0.5mm}\Tr(\mathbf{V})\hspace*{-0.5mm}+\hspace*{-0.5mm}\phi\Lambda^{(i)}\\
\mathrm{s.t.} &&\hspace*{10mm}\mbox{C1 -- C9, C10a, C11, C12, C14},
\end{eqnarray}
where $\Lambda^{(i)}=\Big(\sum_{l=1}^L \sum_{k=1}^K  s_{l,k} \hspace*{-0.5mm}-\hspace*{-0.5mm}\sum_{l=1}^L \sum_{k=1}^K  (s_{l,k}^{(i)})^2 \hspace*{-0.5mm}-\hspace*{-0.5mm}2\sum_{l=1}^L \sum_{k=1}^K s_{l,k}^{(i)} (s_{l,k} \hspace*{-0.5mm}-\hspace*{-0.5mm}s_{l,k}^{(i)})\Big)$,
leads to an upper bound of (\ref{eqn:equivalent-approx2}). Then, an iterative algorithm is used to tighten the upper bound as summarized in Table \ref{table:algorithm2}.  We first initialize the values of $s_{l,k}^{(i)}$ and the iteration index  $i=0$. Then, we solve (\ref{eqn:equivalent-approx3}) for a given value of $s_{l,k}^{(i)}$, cf. line 3. Subsequently, we update $s_{l,k}^{(i+1)}$ with the intermediate solution $s_{l,k}'$.  The main idea of the proposed iterative method is to generate a sequence of feasible solutions $s_{i,k}^{(i)}$ by successively solving the convex upper bound
problem (\ref{eqn:equivalent-approx3}). The procedure is repeated iteratively until convergence  or the maximum number of iterations is reached. We note that the proposed suboptimal algorithm converges to a locally optimal solution of (\ref{eqn:equivalent-approx2}) with polynomial time computational complexity as shown in  \cite{book:SCA_convergence}. Besides, by exploiting  Theorem 1, $\Rank(\mathbf{W}_k)=1$ is guaranteed despite the adopted SDP relaxation. On the contrary, although the optimal resource allocation algorithm achieves
the optimal system performance, it has a non-polynomial time computational complexity.
\begin{Remark}
The proposed algorithm requires  $s_{l,k}^{(i)}$ to be a feasible point  for the initialization, i.e., $i=0$. This point can be obtained by e.g. solving (\ref{eqn:equivalent-approx2}) for $\phi=0$.
\end{Remark}

\begin{Remark}{
The computational complexity of the proposed suboptimal algorithm with respect to the number of IRs $K$, the number of ERs $M$, and the total number of transmit antennas $N_{\mathrm{T}}L$  is given by \cite{book:interior_point}
\begin{eqnarray}
&&\hspace*{-9mm}\bigo\Bigg(
\Big((K\hspace*{-0.5mm}+\hspace*{-0.5mm}MK\hspace*{-0.5mm}+\hspace*{-0.5mm}2L\hspace*{-0.5mm}+\hspace*{-0.5mm}M\hspace*{-0.5mm}+\hspace*{-0.5mm}KL)(2N_{\mathrm{T}}L)^3\\
&&\hspace*{-9mm}\hspace*{-0.5mm}+\hspace*{-0.5mm}(2N_{\mathrm{T}}L)^2
(K\hspace*{-0.5mm}+\hspace*{-0.5mm}MK\hspace*{-0.5mm}+\hspace*{-0.5mm}2L\hspace*{-0.5mm}+\hspace*{-0.5mm}M\hspace*{-0.5mm}+\hspace*{-0.5mm}KL)^2\notag\\
&&\hspace*{-9mm}
\hspace*{-0.5mm}+\hspace*{-0.5mm}(K\hspace*{-0.5mm}+\hspace*{-0.5mm}MK\hspace*{-0.5mm}+\hspace*{-0.5mm}2L\hspace*{-0.5mm}+
\hspace*{-0.5mm}M\hspace*{-0.5mm}+\hspace*{-0.5mm}KL)^3\Big)\hspace*{-0.5mm}T_{\mathrm{Iter}}
\Big(\sqrt{2N_{\mathrm{T}}L}\log(\frac{1}{\Delta})\Big)\Bigg)\notag
\end{eqnarray}
for a given solution accuracy $\Delta>0$ of the adopted numerical solver, where  $\bigo(\cdot)$ is the big-O notation and $T_{\mathrm{Iter}}$ is the number of iterations required for the proposed suboptimal algorithm. We note that the proposed suboptimal algorithm has a polynomial time computational complexity which  is considered to be low, cf. \cite[Chapter 34]{book:polynoimal}, and is desirable for real time implementation. Besides, the computational complexity of the proposed suboptimal algorithm can be further reduced by adopting a tailor made interior point method \cite{JR:complexity1,JR:complexity2}.}
\end{Remark}

\begin{figure}[t]
\centering
\includegraphics[width=3.5in]{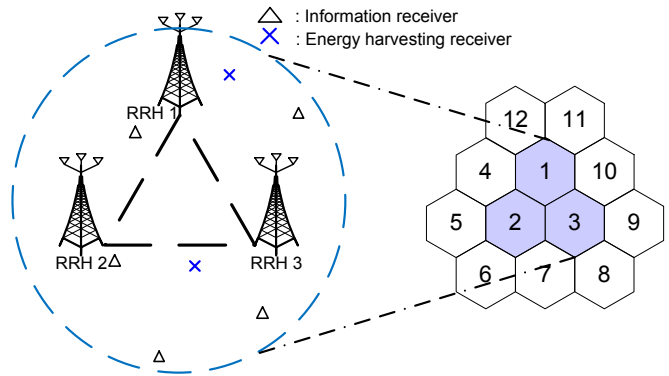}
\caption{Adopted two-tier distributed antenna  network simulation topology. There are $L=3$ cooperative RRHs serving $K=5$ IRs and $M=2$ ERs in the first tier network (shaded area). RRH $4$ -- RRH $12$ are non-cooperative RRHs which serve only the IRs in the second tier (unshaded area).}
\label{fig:simulation_model}
\end{figure}

\begin{table*}[t]\caption{System parameters.}\label{tab:feedback} \centering
\begin{tabular}{|l|l|}\hline
\hspace*{-1mm}Carrier center frequency and path loss exponent & $915$ MHz and  $2.7$ \\
\hline
\hspace*{-1mm}Multipath fading distribution & \mbox{Rayleigh fading}  \\

\hline
\hspace*{-1mm}Thermal and signal processing noise power, $\sigma_{\mathrm{s}}^2$ &  $-23$ dBm   \\
\hline
\hspace*{-1mm}Circuit power consumption at the CP and the $l$-th cooperative RRH &  $40$ dBm and $30$ dBm  \\
\hline
\hspace*{-1mm}Power amplifier  efficiency  & $1/{\rho}=0.38$  \\
\hline
\hspace*{-1mm}Max. transmit power  allowance, $P_l^{T_{\max}}$, and min. required power transfer\footnote{ The minimum required power transfer of $-10$ dBm is suitable e.g. for sensor  applications.} & $48$ dBm and $-10$ dBm \\
\hline

\hspace*{-1mm}RF to electrical energy conversion efficiency, $\mu$, and penalty term, $\phi$ & $0.5$ and $10P_l^{T_{\max}}$  \\
\hline
\hspace*{-1mm}B-coefficient matrix&  Obtained from example 4D in \cite{book:B_matrix}  \\
           \hline

\end{tabular}
\end{table*}
\section{Results}
In this section, we evaluate  the network performance of the proposed resource allocation algorithms via simulations.
{We focus on a two-tier distributed antenna network,  cf. Figure \ref{fig:simulation_model}, which includes the impact of multicell interference on the proposed algorithm design. We assume that  RRH $1$, $2$, and $3$ are connected to the CP,  i.e., $L=3$,  to form a  cooperative cluster for serving a  \emph{heavily loaded area} in a multicell system (shaded area in Figure \ref{fig:simulation_model}).
 There are $K=5$ IRs and $M=2$ ERs in the cooperative cluster. The inter-site distance between any two cooperative RRHs is $150$ meters which is a typical distance for a micro-cellular setup. The three cooperative RRHs form an equilateral triangle while the IRs and ERs are uniformly distributed inside a disc with  radius  $150$ meters centered at the centroid of the triangle.
  The second tier is a \emph{lightly loaded area} served by  RRH $4$ -- RRH $12$ (unshaded area in Figure \ref{fig:simulation_model}). These RRHs are non-cooperative  RRHs each serving the IRs in one of the $9$ cells in the second tier. The distance between two neighboring non-cooperative RRHs is $150$ meters and   each non-cooperative RRH is  located at the center of a second tier cell with cell radius  $75$ meters.  In each second tier cell, one IR is uniformly  and randomly distributed  requiring a minimum SINR of $6$ dB and no communication security. Besides, each  non-cooperative RRH is powered by a non-renewable energy source and equipped with $N_{\mathrm{T}}^{\mathrm{NC}}=5$ transmit antennas. Furthermore,  the  non-cooperative RRHs  do not require the backhaul  for downlink transmission.  The objective of each non-cooperative RRH is to minimize its own transmit power subject to the minimum required   SINR constraint.}   The performance of the proposed algorithms is compared with the performances of a  fully cooperative transmission scheme\footnote{Throughout this section, ``full cooperation" refers to full cooperation in the first tier of the network.} (cooperative transmission and energy cooperation), a  fully cooperative transmission scheme  with perfect CSI but without (w/o) energy cooperation, and a traditional system with co-located transmit antennas. For the fully cooperation scheme, the solution is obtained by setting $C_{l}^{\mathrm{B}_{\max}}\rightarrow\infty$, and solving (\ref{eqn:cross-layer}) by SDP relaxation. For the fully cooperative scheme with perfect CSI but w/o energy cooperation, we set $C_{l}^{\mathrm{B}_{\max}}\rightarrow\infty$ and  $P_l^{T_{\max}}=\infty$, but  restrict the cooperative RRHs to not share the harvested energy, and solve (\ref{eqn:cross-layer}) by SDP relaxation.
   For the co-located transmit antenna system, we assume that there is only one cooperative RRH located  at the center of the cooperative cluster, which is equipped with the same number of antennas as all first tier cooperative RRHs combined in the distributed stetting, i.e., $N_{\mathrm{T}}L$. Besides, for the  co-located transmit antenna system, the CP is at the same location as the  RRH and the backhaul is not needed. Furthermore, we set $P_l^{T_{\max}}=\infty$ and assume an unlimited energy supply for the co-located transmit antenna system to study its power consumption.  Unless specified otherwise, we assume  that the maximum  SINR tolerance of each ER is set to  $\Gamma_{\mathrm{tol}}=0$ dB. We adopt  an Euclidean sphere for the CSI uncertainty region, i.e.,  $\mathbf{\Xi}_m=\mathbf{I}_{N_{\mathrm{T}}L}$. Furthermore, we define the normalized maximum  channel estimation error of ER $m$  as  $\sigma_{\mathrm{est}_m}^2=\frac{\varepsilon^2_m}{\norm{\mathbf{g}_m}^2}=0.05$, where $\forall m\in\{1,\ldots,M\}$. The  parameters adopted in the simulations are summarized in Table \ref{tab:feedback}.

\begin{figure}
\includegraphics[width=3.5 in]{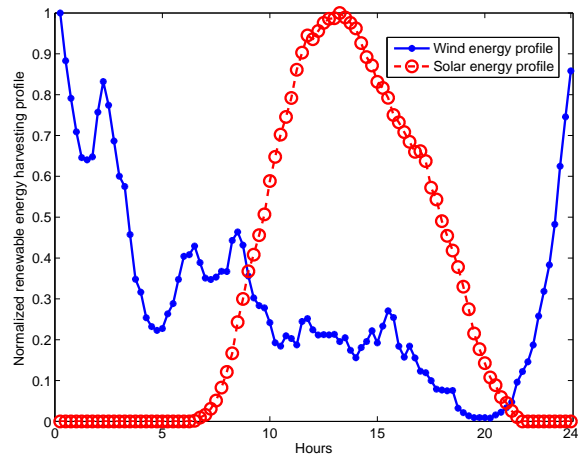}
\caption{ Normalized renewable energy harvesting profile for the considered distributed antenna network. } \label{fig:energy_harvesting_profile} \end{figure}

\begin{figure}
\includegraphics[width=3.5 in]{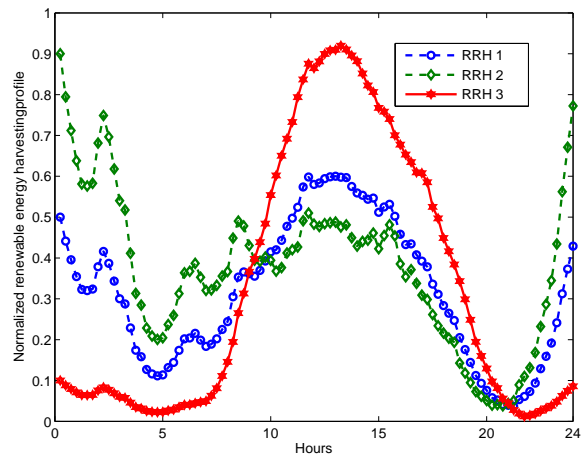}
\caption{Normalized  renewable energy  harvesting profile for the three RRHs. } \label{fig:energy_harvesting_profile_RRH}
 \end{figure}

    Moreover, we adopt the  normalized renewable energy harvesting profile  specified in Figure \ref{fig:energy_harvesting_profile}, for which the data was obtained at August $01, 2014$, in Belgium\footnote{Please refer to http://www.elia.be/en/grid-data/power-generation/ for  details regarding the energy harvesting data.}. The data is averaged  over $15$ minutes, i.e.,  there are $96$ sample points per $24$ hours. We denote the normalized renewable energy harvesting  profile data points for wind energy and solar energy as $\boldsymbol{\xi}_{\mathrm{w}}=[{\xi}_{\mathrm{w},1},\ldots,{\xi}_{\mathrm{w},96}]$ and $\boldsymbol{\xi}_{\mathrm{s}}=[{\xi}_{\mathrm{s},1},\ldots,{\xi}_{\mathrm{s},96}]$, respectively.
We follow a similar approach as in \cite{JR:CoMP_energy} to generate the amount of harvested energy at each cooperative RRH for simulation.  We assume that the CP has only enough energy to support its circuit power consumption  and does not contribute energy to the energy cooperation between the cooperative RRHs. The three cooperative RRHs are equipped with both solar panels and wind turbines with different energy harvesting capabilities. The harvested energy over time at the three cooperative RRHs is given by $\boldsymbol{\xi}_1=E (0.5 \boldsymbol{\xi}_{\mathrm{w}}+0.5\boldsymbol{\xi}_{\mathrm{s}})$, $\boldsymbol{\xi}_2=E (0.9 \boldsymbol{\xi}_{\mathrm{w}}+0.1\boldsymbol{\xi}_{\mathrm{s}})$, and $\boldsymbol{\xi}_3= E(0.1 \boldsymbol{\xi}_{\mathrm{w}}+0.9\boldsymbol{\xi}_{\mathrm{s}})$, respectively, as shown in Figure \ref{fig:energy_harvesting_profile_RRH}, where  $E=500$ Joules  is a given constant indicating the maximum available energy from the solar panels and wind turbines.  Thus, the maximum harvested energy  for cooperative RRH $n\in\{1,\ldots,L\}$ at sample time $r\in\{1,\ldots,96\}$ is given by  $E^{\max}_n= \big[\boldsymbol{\xi}_n\big]_{1,r}$.  The minimum required received SINRs for the five IRs are set to $\Gamma_{\mathrm{req}_k}=[ 6,\, 9, \,12, \,15,\,18]$ dB, respectively. In case of first tier full cooperation, these five IRs require a total capacity of $15.5818$ bit/s/Hz which is the aggregated secrecy rate of all IRs, i.e.,   $\sum_{k=1}^K(\log_2(1+\Gamma_{\mathrm{req}_k})-\log_2(1+\Gamma_{\mathrm{tol}}))$.


\subsection{Convergence of the Proposed Iterative Algorithms}
\begin{figure}[t]
\centering
\includegraphics[width=3.5in]{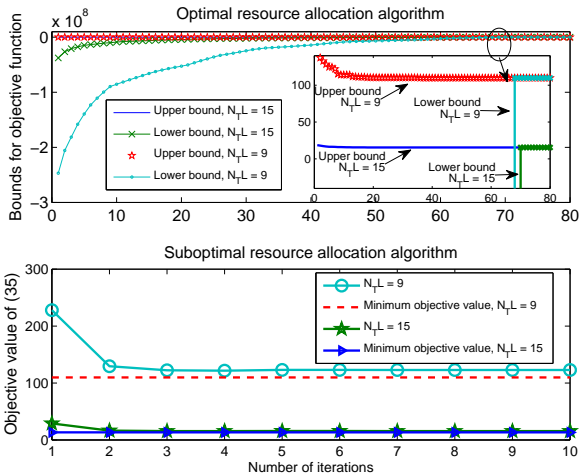}
\caption{Convergence of the proposed iterative algorithms. }
\label{fig:convergence}
\end{figure}

Figure \ref{fig:convergence} illustrates the convergence of
the proposed optimal and suboptimal algorithms  for  different total numbers of transmit antennas in the network, $N_{\mathrm{T}}L$. The backhaul capacity per link is $10$ bits/s/Hz. It can be seen from the upper half of Figure \ref{fig:convergence} that
the proposed optimal algorithm converges to the optimal solution, i.e., the upper bound value meets the lower bound value after less than $80$ iterations. On the other hand, the suboptimal algorithm converges to a locally optimal value after less than $10$ iterations.  We note that if a brute force approach is adopted to obtain a global optimal solution without exploiting the structure of the problem, for $K=5$ IRs and $L=3$ cooperative RRHs,
$2^{15}$ of SDPs need to be solved  which may not be computational feasible in practice.

\begin{figure}[t]
 \centering
\includegraphics[width=3.5 in]{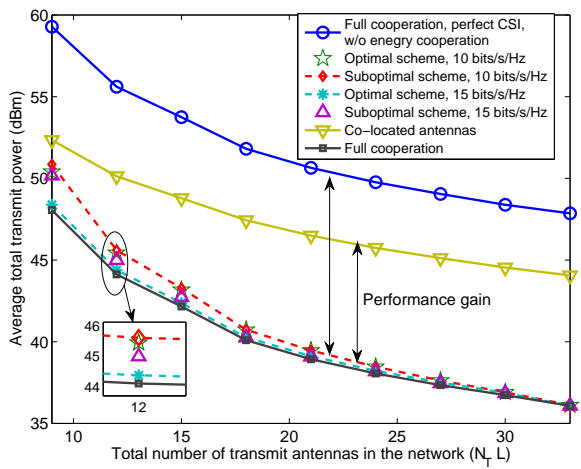}
\caption{Average total transmit power (dBm) versus the total number of transmit antennas in the network, $N_{\mathrm{T}}L$.} \label{fig:tp_nt}
\end{figure}
\begin{figure}[t]
\includegraphics[width=3.5 in]{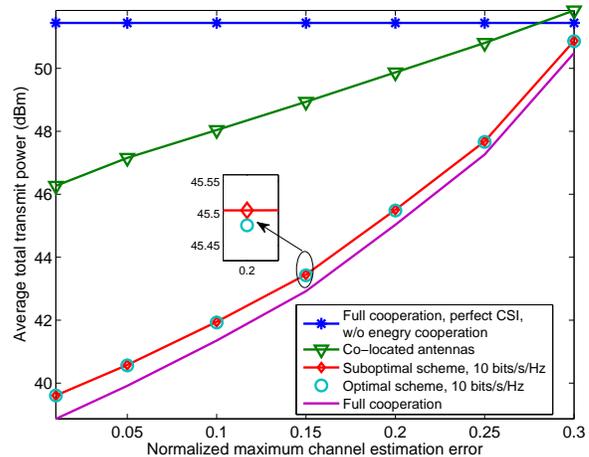}
\caption{Average total transmit power (dBm)  versus the normalized channel estimation error for different  resource allocation schemes. } \label{fig:pt_errors}
\end{figure}

\subsection{Average Total Transmit Power}
In Figure \ref{fig:tp_nt}, we study the average total transmit power versus the total numbers of transmit antennas in the network, $N_{\mathrm{T}}L$, for different resource allocation schemes. The performances of the proposed optimal and suboptimal iterative algorithms are shown for $80$ and $10$ iterations, respectively.  It can be seen  that the transmit power for the proposed optimal and suboptimal schemes decreases when the backhaul capacity per backhaul link increases from $10$ bits/s/Hz to $15$ bits/s/Hz. This is because the increased backhaul capacity facilitates joint transmission and thus reduces the total transmit power. However, the transmit power of all considered schemes/systems decreases gradually with the  total number of transmit antennas in the network. In fact,  extra degrees of freedom can be exploited for resource allocation when  more antennas are available for the cooperation between the RRHs.  Furthermore, the performance gap between the proposed optimal algorithm and fully cooperative transmission   is expected to decrease with increasing $N_{\mathrm{T}}L$. For sufficiently large  numbers of antennas at the cooperative RRHs, conveying the data of each IR to a subset of cooperative RRHs via the backhaul links may be sufficient for guaranteeing the QoS requirements for reliable communication and efficient power transfer. The lower average total transmit power of fully cooperative transmission with energy cooperation  comes at the expense of an exceedingly high backhaul capacity consumption. On the other hand, the proposed suboptimal algorithm achieves an excellent system performance even for the case of only $10$ iterations.

Compared to the two proposed schemes, it is expected that the co-located antenna scheme requires a higher transmit power since the co-located antenna system does not offer
network type spatial diversity  to combat the path loss. Furthermore, Figure \ref{fig:tp_nt} reveals that the performance of fully cooperative transmission with perfect CSI and w/o energy cooperation is significantly worse than that of all other schemes. Specifically, cooperative RRH $3$ mainly relies on the solar panel for energy harvesting and thus the available energy for cooperative RRH $3$ is very limited during the night time. Therefore, despite the availability of  perfect CSI and a large number of distributed antennas in the system,  the cooperative RRHs having more
harvested renewable
energy available are required to transmit  with comparatively large powers for assisting the cooperative RRHs with smaller harvested renewable
energy.  In fact, the cooperative RRHs have to cooperate wirelessly which is less power efficient than the cooperation via the micro-grid.

In Figure \ref{fig:pt_errors}, we show the average total transmit power (dBm)  versus the normalized channel estimation error for the proposed schemes with $N_{\mathrm{T}}L=18$ and $10$ bits/s/Hz capacity per backhaul link. As can be observed, the average transmit power increases with the normalized channel estimation error except for the case of perfect CSI.  The reason behind this is twofold. First, a higher transmit power for the artificial noise, $\mathbf{v}$,  is required to satisfy constraints C2 and C7 due to a larger uncertainty set for the CSI, i.e., $\mathbf{\Xi}_m$. Second, a higher amount of power also has to be allocated to the information signal $\mathbf{w}_ks_k,\forall k$, cf. $\mathbf{w}_ks_k,\forall k$,  for neutralizing the interference caused by the artificial noise at the desired IRs.

\begin{figure}[t]
 \centering
\includegraphics[width=3.5 in]{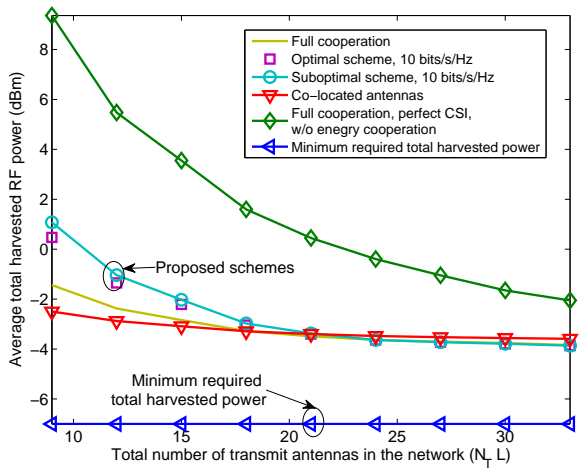}
\caption{Average total harvested RF power (dBm) versus the  the total number of transmit antennas in the network, $N_{\mathrm{T}}L$, for different  resource allocation schemes. } \label{fig:hp_nt}
\end{figure}
\begin{figure}[t]
 \includegraphics[width=3.5 in]{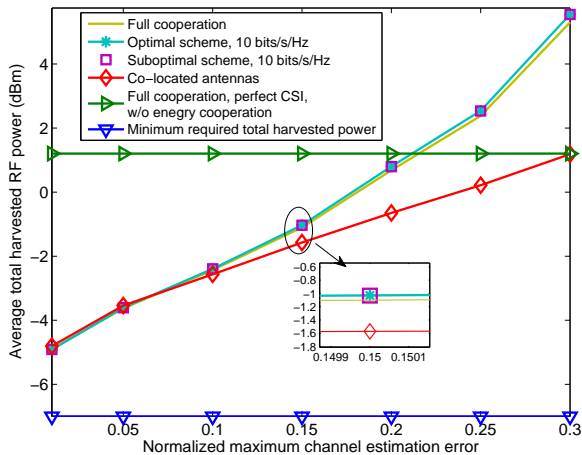}
\caption{Average total harvested RF power (dBm) versus the normalized channel estimation error for different  resource allocation schemes.} \label{hp_errors}
\end{figure}

\subsection{Average Total Harvested Power}
In Figure \ref{fig:hp_nt}, we study the average total harvested RF power  versus the total number of transmit antennas for  different resource allocation schemes.  It can be observed that the total harvested power of the proposed schemes decreases monotonically with increasing  number of transmit antennas. This is because the extra degrees of freedom offered by the increasing number of antennas improve the efficiency of resource allocation. In particular,  the direction of beamforming
matrix $\mathbf{W}_k$ can be more accurately steered towards the IRs which reduces the power allocation to $\mathbf{W}_k$ and the leakage of power to the ERs.  This also explains the lower harvested power for fully cooperative transmission with energy cooperation which can exploit all transmit antennas in the network for joint transmission. Besides, for the fully cooperative  scheme w/o energy cooperation, the ERs harvest the highest amount of power on average at the expense of the highest average total transmit power. Furthermore, although the system with co-located antennas consumes a  higher transmit power, it  does not always lead to the largest harvested power at the ERs in all considered scenarios. Indeed, a large portion of radiated power in the co-located antenna system is used to combat the path loss which emphasizes the benefits of the inherent spatial diversity in distributed antenna systems for power efficient transmission. We also show in Figure \ref{fig:hp_nt}  the minimum required total harvested power which is computed  by assuming that constraint C7 is satisfied  with equality for all ERs.  Despite the imperfection of the CSI, because of the adopted robust optimization framework, the proposed  optimal  and   suboptimal resource allocation schemes are able to guarantee the minimum harvested energy required by constraint C7 in every time instant. On the other hand, Figure \ref{hp_errors} depicts the average total harvested power versus the normalized
channel estimation error
 for the proposed schemes with $N_{\mathrm{T}}L=18$ and $10$ bits/s/Hz backhaul capacity per backhaul link. For imperfect CSI, the harvested power increases with the channel estimation error. In fact,  to fulfill  the QoS requirements on power transfer and communication secrecy, more transmit power is required  for larger $\sigma_{\mathrm{est}_m}^2$ which leads to a higher energy level  in the RF for energy harvesting.

 \begin{Remark}{
 We note that for all scenarios considered in this section, the proposed
resource allocation schemes are able to guarantee the required secrecy rate for all IRs, i.e.,  $R_{\mathrm{sec}_k}= \log_2(1+\Gamma_{\mathrm{req}_k})-\log_2(1+\Gamma_{\mathrm{tol}})$, despite the imperfectness of the CSI of the ERs.}
 \end{Remark}

\section{Conclusions}\label{sect:conclusion}
In this paper, we studied the resource allocation algorithm design for the wireless delivery of both secure information and renewable green energy to mobile receivers in  distributed antenna  communication systems.  The algorithm design was formulated as a non-convex optimization problem with the objective to  minimize the total network transmit power. The proposed problem formulation took into account the limited backhaul capacity, the sharing of harvested renewable green energy between RRHs,  the  imperfect CSI of the ERs, and QoS requirements for secure communication and efficient  power transfer.  An optimal  iterative resource allocation algorithm was proposed for obtaining a global optimal solution based on the generalized Bender's decomposition. To strike a balance between computational complexity and optimality, we also proposed a low complexity suboptimal algorithm. Simulation results showed that the proposed suboptimal iterative resource allocation scheme  performs close to the optimal  scheme. Besides, our results unveiled the potential power savings in SWIPT systems employing distributed antenna  networks and renewable green energy sharing compared to centralized systems with multiple co-located antennas.
\section*{Appendix-Proof of Proposition \ref{thm:supporting_plane}}
We start the proof by studying the solution of the dual problem in (\ref{eqn:dual}). For a given optimal dual variable  $\boldsymbol \Theta(i)$, we have $\boldsymbol \Theta(i)$
\begin{eqnarray}\notag
\notag\hspace*{-3.5mm}&=&\hspace*{-2.5mm}\arg\underset{{\mathbf{W}_k,\mathbf{V}\in\mathbb{H}^{N_{\mathrm{T}}L},\mathbf{e}^{\mathrm{S}},\boldsymbol \delta, \boldsymbol \nu }}{\min}\,{\cal L}\Big(\mathbf{W}_k,\mathbf{V},\mathbf{e}^{\mathrm{S}},\boldsymbol \delta, \boldsymbol \nu,  s_{l,k}(i),\boldsymbol \Phi(i)\Big)\\
\notag\hspace*{-3.5mm}&=&\hspace*{-2.5mm} \arg \underset{\underset{\mathbf{e}^{\mathrm{S}},\boldsymbol \delta, \boldsymbol \nu }{\mathbf{W}_k,\mathbf{V}\in\mathbb{H}^{N_{\mathrm{T}}L}}}{\min}\,f_0(\mathbf{W}_k,\mathbf{V})\hspace*{-0.5mm}+\hspace*{-0.5mm}
f_1(\mathbf{W}_k,\mathbf{V},\mathbf{e}^{\mathrm{S}},\boldsymbol \delta, \boldsymbol \nu,  \boldsymbol \Phi(i)) \hspace*{-0.5mm}\notag\\ \notag\hspace*{-3.5mm}&=&\hspace*{-2.5mm} \sum_{k=1}^K\hspace*{-0.5mm}\sum_{l=1}^L\hspace*{-0.5mm}\beta_{k,l}\Big(\hspace*{-0.5mm}\Tr(\mathbf{W}_k\mathbf{R}_l)\hspace*{-0.5mm}-
\hspace*{-0.5mm} s_{l,k}(i) P^{\mathrm{T}_{\max}}_l\hspace*{-1mm}\Big)\\
\hspace*{-3.5mm}&=&\hspace*{-2.5mm} \arg\underset{\underset{\mathbf{e}^{\mathrm{S}},\boldsymbol \delta, \boldsymbol \nu }{\mathbf{W}_k,\mathbf{V}\in\mathbb{H}^{N_{\mathrm{T}}L}}}{\min}\,
f_0(\mathbf{W}_k,\mathbf{V})\hspace*{-0.5mm}+\hspace*{-0.5mm}f_1(\mathbf{W}_k,\mathbf{V},\mathbf{e}^{\mathrm{S}},\boldsymbol \delta, \boldsymbol \nu,  \boldsymbol \Phi(i)) \notag\\ \hspace*{-3.5mm}&+&\hspace*{-2.5mm} \sum_{k=1}^K\hspace*{-0.5mm}\sum_{l=1}^L\hspace*{-0.5mm}\beta_{k,l}\Tr(\mathbf{W}_k\mathbf{R}_l),\label{eqn:supporting_plane_eq1}
\end{eqnarray}
where the first equality is due to the  Karush-Kuhn-Tucker (KKT) conditions of the SDP relaxed problem in (\ref{eqn:equivalent}). On the other hand, we can rewrite function $\xi( \boldsymbol{\Phi}(t),s_{l,k}),t\in\{1,\ldots,i\}$ as
\begin{eqnarray}\notag
&&\xi( \boldsymbol{\Phi}(t),s_{l,k})\notag\\
\hspace*{-1.5mm}&=&\hspace*{-1.5mm} \underset{\underset{\mathbf{e}^{\mathrm{S}},\boldsymbol \delta, \boldsymbol \nu }{\mathbf{W}_k,\mathbf{V}\in\mathbb{H}^{N_{\mathrm{T}}L}}}{\mino}\,\, \Big\{f_0(\mathbf{W}_k,\mathbf{V})+f_1(\mathbf{W}_k,\mathbf{V},\mathbf{e}^{\mathrm{S}},\boldsymbol \delta, \boldsymbol \nu,  \boldsymbol \Phi(t)) \notag\\
\hspace*{-1.5mm}&+&\hspace*{-1.5mm} f_2(\mathbf{W}_k, s_{l,k},\boldsymbol \Phi(t))\Big\}\\
\hspace*{-1.5mm}&=&\hspace*{-1.5mm}\Bigg\{\underset{\underset{\mathbf{e}^{\mathrm{S}},\boldsymbol \delta, \boldsymbol \nu }{\mathbf{W}_k,\mathbf{V}\in\mathbb{H}^{N_{\mathrm{T}}L}}}{\mino}\,f_0(\mathbf{W}_k,\mathbf{V})\hspace*{-0.5mm}
+\hspace*{-0.5mm}f_1(\mathbf{W}_k,\mathbf{V},\mathbf{e}^{\mathrm{S}},\boldsymbol \delta, \boldsymbol \nu,  \boldsymbol \Phi(t))\notag\\
\hspace*{-1.5mm}&+&\hspace*{-1.5mm}\sum_{k=1}^K\sum_{l=1}^L\beta_{k,l}\Tr(\mathbf{W}_k\mathbf{R}_l)\Bigg\}\notag\\
\hspace*{-1.5mm}&-&\hspace*{-1.5mm}\sum_{k=1}^K\sum_{l=1}^L\beta_{k,l} s_{l,k} P^{\mathrm{T}_{\max}}_l. \label{eqn:supporting_plane_eq2}
\end{eqnarray}
As a result,  the primal solution in the $t$-th iteration, $\boldsymbol \Theta(t)$, is also the solution for the minimization in the master problem in (\ref{eqn:supporting_plane_eq2}) for the $t$-th constraint in (\ref{eqn:supporting_plane_constraint1}). Similarly, we can  use the same approach to prove that the solution of (\ref{eqn:FP}) is also the solution of (\ref{eqn:opt_master_2}). \qed


\end{document}